\documentclass[3p,review,11pt]{elsarticle}
\usepackage{amssymb}
\usepackage{amsmath}
\usepackage{theorem}

\usepackage{graphicx}
\usepackage{multirow}
\usepackage{subfigure}

\usepackage{multicol}
\usepackage[ruled,noline,linesnumbered]{algorithm2e}
\usepackage{caption}
\usepackage{lscape}
\usepackage[T1]{fontenc}
\usepackage{algorithmic}
\usepackage{float}
\usepackage{rotating}
\usepackage[usenames,dvipsnames]{xcolor}
\usepackage{lineno}
\usepackage{tabularx}
\usepackage{url}

\newcommand{\rememberlines}{\xdef\rememberedlines{\number\value{AlgoLine}}}

\let\oldnl\nl
\newcommand{\nonl}{\renewcommand{\nl}{\let\nl\oldnl}}

{\theorembodyfont{\rmfamily}

}

{\theorembodyfont{\rmfamily}

}
{\theorembodyfont{\rmfamily}

}
{\theorembodyfont{\rmfamily}

}

\begin{document}

\title{Optimization in Sanger sequencing\tnoteref{1}}\tnotetext[1]{Link to published version: \url{https://doi.org/10.1016/j.cor.2019.05.011}. Under the CC-BY-NC-ND 4.0 license (\url{https://creativecommons.org/licenses/by-nc-nd/4.0/}).}

\author[mt]{Luisa Carpente}
\ead{luisa.carpente@udc.es}

\author[bd]{Ana Cerdeira-Pena}
\ead{acerdeira@udc.es}

\author[mt]{Silvia Lorenzo-Freire\corref{cor1}}
\ead{silvia.lorenzo@udc.es}

\author[bd]{\'Angeles S. Places}
\ead{asplaces@udc.es}

\cortext[cor1]{Corresponding author. Postal code: 15071, A Coru\~{n}a. Phone: +34 981167000. Fax: +34 981167160.}
\address[bd]{Database Laboratory, Department of Computer Science, Faculty of Computer Science, University of A Coru\~{n}a, Spain}
\address[mt]{MODES Research Group, Department of Mathematics, Faculty of Computer Science, University of A Coru\~{n}a, Spain}
\begin{abstract}
The main objective of this paper is to solve the optimization problem that is associated with the classification of DNA samples in PCR plates for Sanger sequencing. To achieve this goal, we design an integer linear programming model. Given that the real instances involve the classification of thousands of samples and the linear model can only be solved for small instances, the paper includes a heuristic to cope with bigger problems.

The heuristic algorithm is based on the simulated annealing technique. This algorithm obtains satisfactory solutions to the problem in a short amount of time. It has been tested with real data and yields improved results compared to some commercial software typically used in (clinical) laboratories. Moreover, the algorithm has already been implemented in the laboratory and is being successfully used.
\end{abstract}

\maketitle

\section{Introduction}
Recent developments in the study of DNA will have a substantial impact on various fields of science, such as clinical medicine and forensic studies. Increasingly many companies and laboratories are processing DNA samples for various purposes and competing in terms of the time and cost that are required for processing the results. Because optimization techniques can be used to reduce the time or cost of any process, these techniques are of great interest for DNA sample processing.

In this article, we will address a scheduling problem that arises within Health in Code\footnote{\url{http://www.healthincode.com}}, which is a company that specializes in the genetic diagnosis of inherited cardiovascular diseases. Every day, thousands of DNA samples are processed in their labs.

DNA sequencing is the process that is used to determine the order of the four nucleotides, namely, Adenine (A), Cytosine (C), Guanine (G) and Thymine (T), that make up the DNA molecule. Currently, there are several methods for DNA sequencing. Health in Code is a company that has extensive experience with the Sanger method \cite{sanger77}. The method, which is also referred to as dideoxynucleotide sequencing and chain termination sequencing, consists of the following steps:

\begin{itemize}
\item DNA extraction (from samples of tissue, saliva, blood, etc.) and dilution treatment;
\item Polymerase chain reaction (PCR): an effective process for making copies of segments of DNA;
\item PCR product purification: to remove elements that are used in the PCR process to obtain high-quality DNA samples for sequencing;
\item DNA sequencing: to sort the DNA fragments by size in a sequencing machine so that the original piece of DNA can be decoded.
\end{itemize}

Although all these steps are essential in the Sanger method, this paper focuses on the PCR process. The polymerase chain reaction is a method for amplifying DNA to generate millions of copies of one or several pieces of DNA. To perform this reaction, the DNA is deposited into PCR plates. Then, the plates are placed into thermocyclers\footnote{A thermocycler, which is also known as a thermal cycler, is a laboratory machine that allows several temperatures to be set in a block of a plate.} to regulate temperature during cyclical programmes.

In this framework, we are interested in optimizing the organization of DNA samples in the plates for performing the PCR process, which is not an easy task since many aspects must be considered to achieve the optimal organization of DNA samples. Each DNA sample is deposited into the well of a plate; however, depending on the components that are used in the procedure, each well from a PCR plate can be used to obtain copies of a single piece of DNA or copies of a group of pieces of DNA. The components that are used with the DNA sample determine not only the piece or pieces to be copied but also the necessary temperature for processing the well in the thermocycler. However, due to the technological characteristics of thermocyclers, wells in the same plate should satisfy several conditions. According to their characteristics, each plate is divided into several strips and all the wells in the same strip should be processed at the same temperature. Moreover, the temperatures of the consecutive strips in the same plate should belong to a specified range.

Since processing each plate in the laboratory is slow and expensive, the main objective of our optimization problem is to order the DNA samples such that the number of plates is minimized under the restrictions that are imposed by the characteristics of the thermocycler. It is also desirable to obtain plates that are as full as possible.

To the best of our knowledge, the optimization literature has not addressed this topic. Characteristics of the problem remind us of a bin packing problem in which samples must be packed in the plates. The bin packing problem is a well-known NP-hard optimization problem (see, for instance, Delorme et al. \cite{delorme16}). For this reason, it is not always possible to obtain the optimal solution of this problem in a reasonable computational time. Some of the heuristic proposals for solving the bin packing problem use strategies that minimize the slack of each bin (see, for instance, Fleszar and Hindi \cite{fleszar02}). However, in our setting, if we consider various temperature ranges and the constraints that are associated with the reagent that is necessary for amplifying a specified region of DNA, we will be dealing with a different optimization problem in the end.

Due to the characteristics of our problem, although we have developed an ILP model to solve this problem, we have also implemented a heuristic algorithm based on the simulated annealing procedure that provides satisfactory solutions in short computational times. This algorithm is being successfully used in the laboratory.

The remainder of the paper is organized as follows: In Section 2, the problem is described in detail. Section 3 addresses both the optimization model and the heuristic algorithm. In Section 4, we provide a detailed explanation of the numerical experiments that are necessary for justifying an interest in the results. For this purpose, we have considered real instances from the laboratory.

\section{Problem}

\begin{figure}[htbp]
    \centering
    \subfigure[empty plate]{\includegraphics[scale=0.35]{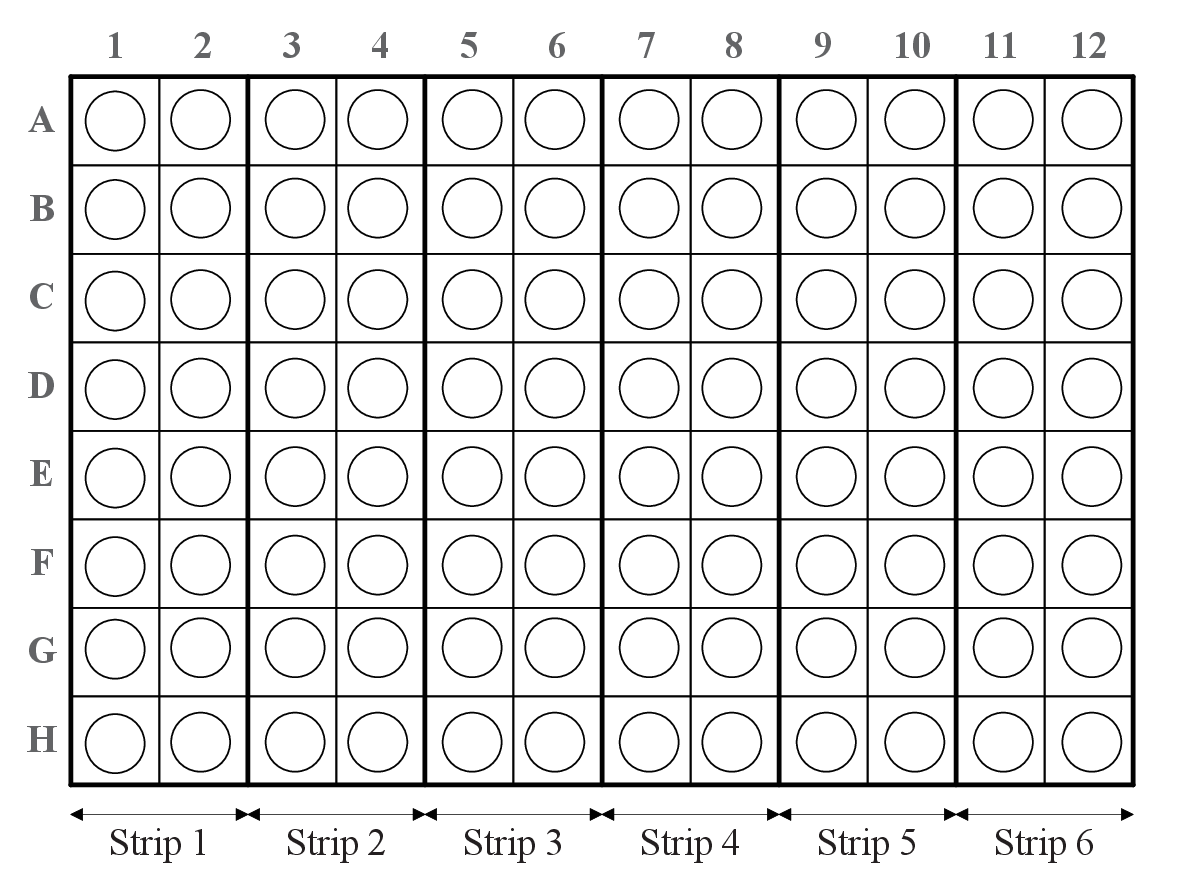}}
    \subfigure[plate with DNA samples]{\includegraphics[scale=0.35]{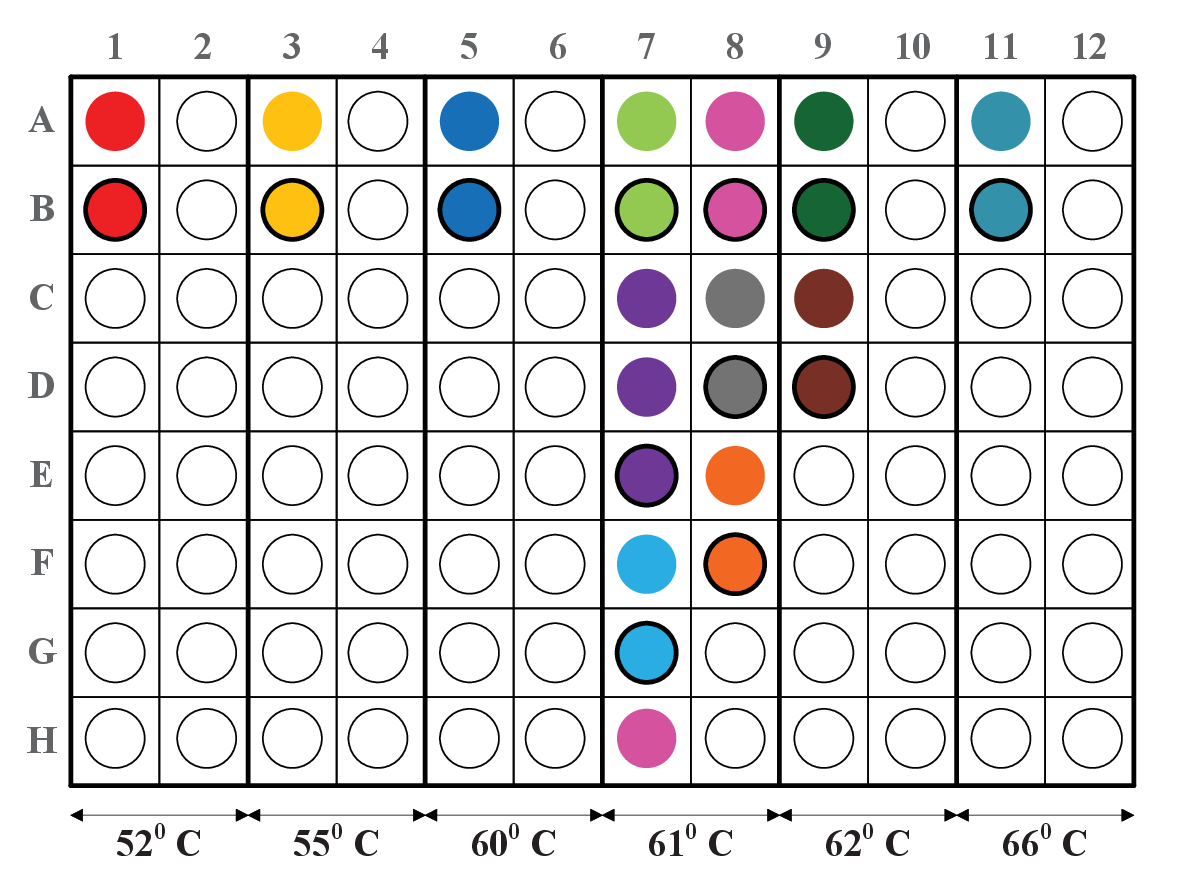}}
    \caption{Example of a PCR plate} \label{fig:oneplate}
\end{figure}

In this problem, the available DNA samples are organized in PCR plates to be processed in thermocyclers. To perform this task properly, several details must be considered:

The laboratory uses 96-well PCR plates and the wells are organized in 8 rows and 12 columns. Each plate is composed of six \textit{strips}. Each strip has 24 wells, which are arranged in 8 rows and 2 columns. According to the characteristics of the thermocyclers, all the wells in the same strip will be processed at the same temperature. In addition, the difference in temperature between two consecutive strips cannot exceed 5 degrees centigrade. Although the strips in a plate must satisfy these temperature conditions, the temperature in the strips is not fixed, but rather is selected by the user. The configuration of a 96-well PCR plate is shown in Figure \ref{fig:oneplate} (a).

Each DNA sample, together with all the chemical components that are necessary for amplifying a specified region of DNA, are deposited in a well. If we are interested in amplifying the same region of DNA for multiple DNA samples, we must include the same chemical components in their corresponding wells. In this case, we say that these DNA samples belong to the same \textit{group}. All DNA samples in a group must be processed at the same temperature, which is fixed. However, it is possible to have multiple groups with the same temperature.

For instance, let us suppose that the laboratory is interested in studying the risk of genetic diseases. In this particular case, we can think of a group as being associated with a disease. Thus, the laboratory must determine, using the DNA samples of the clients, whether the clients will develop this disease. However, since the disease will be detected by evaluating a specified region of DNA, each DNA sample will be mixed with a reagent and processed at the necessary temperature for amplifying this region.

In addition, one well of the plate should be reserved for the isolated reagent that is associated with a group, which is necessary to verify that the experiment is performed correctly.

Figure \ref{fig:oneplate} (b) shows an example of one filled plate. The 12 groups are represented by colour. We know the number of DNA samples in each group and the temperatures that correspond to the groups. There are six temperatures for the groups. With this information, we can assign the temperatures in increasing order to the 6 strips of the plate because the temperature difference between any two consecutive strips does not exceed $5^{0}$ C. With this configuration, we have 6 groups that have the same temperature in Strip 4. Moreover, for each group, we have added the reagent, which is represented by a circle of the same colour (but with a black border), as the group samples; see, for instance, the \textquotedblleft red\textquotedblright\textsf{ }group, which is placed in two wells, one of which contains the DNA sample mixed with the reagent, whereas the other contains only the reagent.

In summary, the laboratory initially has multiple samples, each of which has its own code.  The samples belong to groups that are determined by the region of DNA that will be amplified. The laboratory also knows the temperature  associated with each group. After solving the optimization problem with this information and the restrictions that are imposed by the characteristics of the PCR plates, we obtain the number of samples of each group in every strip of a plate and whether the reagent of each group is present in any strip. We will also know the temperature of every strip, which will be the only information provided by the optimization problem since in the next phase, the software will place each sample or reagent at random but according to this distribution. Then, for each DNA sample or reagent, we will know the code of the plate and the position where it will be placed, the type of analysis (the group to which it belongs) and the temperature at which it will be processed.

\section{Solution methods}
\subsection{Model}
We have designed an integer linear programming (ILP) problem that takes into account all the requirements that were discussed. The main objective of this problem is to minimize the number of occupied plates. We know the number of samples and the temperature for each group. With this information, the ILP problem will assign the samples and the temperature in each strip. Below, we list the parameters and the decision variables that are used in the model.
\begin{center}
\emph{Parameters:}
\end{center}

\begin{itemize}
\item $\{1,\ldots,p\}$ is the set of available plates, with index $q$. Because each plate is divided into six strips, the set of strips can be represented by $\{1,\ldots,6 \cdot p\}$, with index $l$.
\item $w_q$ is the weight  assigned to each plate $q\in \{1,\ldots,p\}$ in the objective function.
\item $\{1,\ldots,n\}$  is the set of groups with index $i$.
\item $N_i$ indicates the number of samples in each group $i\in \{1,\ldots,n\}$.
\item $T_i$ is the temperature used in the process for each group $i\in \{1,\ldots,n\}$. Several groups can be processed at the same temperature. Then, we will consider $\{T^{1},\ldots,T^{m}\}\subseteq \{T_{1},\ldots,T_{n}\}$ as the set of temperatures that are used in the process.
\end{itemize}

\begin{center}
\emph{Decision variables:}
\end{center}
For each strip $l\in \{1,\ldots,6 \cdot p\}$, we define the following variables:
\begin{itemize}
\item $t_l$ is the temperature  selected for $l$.

\item For each group $i\in \{1,\ldots,n\}$,

\begin{itemize}
\item[$\star$] $n_{il}$ represents the number of samples of group $i$ that are placed in strip $l$.

\item[$\star$] $x_{il}$ is a binary variable such that

$x_{il}=\left\{
  \begin{array}{ll}
    1, & \hbox{if the reagent of group $i$ is in strip $l$;} \\
    0, & \hbox{otherwise.}
  \end{array}
\right.$
\end{itemize}

\item For each $j\in \{1,\ldots,m\}$,

\begin{itemize}
\item[$\star$] $y_{jl}$ is a binary variable such that

$y_{jl}=\left\{
  \begin{array}{ll}
    1, & \hbox{if strip $l$ has been assigned temperature $T^{j}$;} \\
    0, & \hbox{otherwise.}
  \end{array}
\right.$
\end{itemize}
\end{itemize}

\begin{center}
\textit{Objective function and constraints:}
\end{center}

{\allowdisplaybreaks
\begin{alignat}{2}
\label{objective}
&\textrm{min }z={\displaystyle\sum_{i=1}^n}\quad{\displaystyle\sum_{q=1}^p}\quad w_q\quad{\displaystyle\sum_{l=6\cdot (q-1)+1}^{6 \cdot q}}n_{il} &&\\
\intertext{subject to}
\label{constraint1}
&{\displaystyle\sum_{i=1}^n}\quad \left(n_{il}+x_{il}\right)\le 16 &  &\forall l\in \{1,\ldots,6 \cdot p\}\\
\label{constraint2}
&t_l-t_{l+1}\le 5 & & \forall l\in \{6\cdot (q-1)+1,\ldots,6 \cdot q-1\}, \forall q\in \{1,\ldots,p\}\\
\label{constraint3}
&t_{l+1}-t_{l}\le 5 & & \forall l\in \{6\cdot (q-1)+1,\ldots,6 \cdot q-1\}, \forall q\in \{1,\ldots,p\}\\
\label{constraint4}
&{\displaystyle\sum_{j=1}^m}y_{jl}\cdot T^{j}= t_l & &\forall l\in \{1,\ldots,6 \cdot p\}\\
\label{constraint5}
&{\displaystyle\sum_{j=1}^m}y_{jl}\le 1 & &\forall l\in \{1,\ldots,6 \cdot p\}\\
\label{constraint6}
&y_{jl}\ge \dfrac{{\displaystyle\sum_{i: T_{i}=T^{j}}}n_{il}}{{\displaystyle\sum_{i=1}^n}N_i} & &\forall l\in \{1,\ldots,6 \cdot p\}, \forall j\in \{1,\ldots,m\}\\
\label{constraint7}
&{\displaystyle\sum_{l=1}^{6\cdot p}}\quad n_{il}=N_i & &\forall i\in \{1,\ldots,n\}\\
\label{constraint8}
&n_{il}-x_{il}\ge 0 & &\forall l\in \{1,\ldots,6 \cdot p\}, \forall i\in \{1,\ldots,n\}\\
\nonumber &{\displaystyle\sum_{l'=6\cdot (q-1)+1}^{6 \cdot q}}N_{i}x_{il'}-n_{il}\ge 0& & \forall i\in \{1,\ldots,n\}, \forall l\in \{6\cdot (q-1)+1,\ldots,6 \cdot q\}, \\
\label{constraint9}
& & & \forall q\in \{1,\ldots,p\}\\
\label{constraint10}
&{\displaystyle\sum_{l=6\cdot (q-1)+1}^{6 \cdot q}}x_{il}\le 1&  &\forall i\in \{1,\ldots,n\}, \forall q\in \{1,\ldots,p\}
\end{alignat}
}

The objective function (\ref{objective}) minimizes the number of occupied plates. To attain this objective, the number of samples in each plate is multiplied by a weight. Although the laboratory is mainly interested in minimizing the number of plates, it would also be desirable to minimize the total number of occupied cells\footnote{The total number of occupied cells is composed of the number of cells that are occupied by all the samples, which is fixed, and the number of cells that are necessary for placing the corresponding reagents, which is variable.} and maximize, according to the lexicographical order\footnote{The laboratory prefers to have the first plates occupied to the detriment of the last ones to be able to place new samples in the last plates while the first ones are being processed. Note that those new samples are allocated once the solution has been obtained and already the first plates have been processed.}, the occupancy rate of the plates. The three objectives can be accomplished by penalizing the plates in the last positions using the weights. Then, the weights will be considered in increasing order to fill in the cells of the first plates. In practice, we use $w_{q}=q$ for all $q\in \{1,\ldots,p\}$.

Constraint (\ref{constraint1}) sets the number of available cells in each strip. Constraints (\ref{constraint2}) and (\ref{constraint3}) guarantee that the absolute value of the difference in temperature between two consecutive strips does not exceed 5 degrees centigrade. Constraints (\ref{constraint4}), (\ref{constraint5}) and (\ref{constraint6}) ensure that the temperature of each strip is the same as the temperature of the groups with samples in the strip. Therefore, if ${\displaystyle\sum_{j=1}^m}y_{jl}=0$, strip $l$ is an empty strip and will not be processed. Otherwise, ${\displaystyle\sum_{j=1}^m}y_{jl}=1$ and all the cells in strip $l$ will be processed at the same temperature. Constraint (\ref{constraint7}) is used to check that all samples are in the cells of the plates, whereas (\ref{constraint8}), (\ref{constraint9}) and (\ref{constraint10}) are associated with the reagent. These last three conditions require that if one or more samples of group $i$ is in plate $q$, the isolated reagent  related to this group should be deposited in exactly one well in the plate; otherwise, ${\displaystyle\sum_{l=6\cdot (q-1)+1}^{6 \cdot q}}x_{il}=0$ implies that there are no samples of group $i$ in plate $q$ and the corresponding reagent will not appear in the plate.

\subsection{Heuristic algorithm}
Because it is necessary to provide feasible and reasonable solutions to the problem in a short computational time, we have designed a heuristic algorithm. The heuristic algorithm is based on the simulated annealing method, which was introduced by Kirkpatrick et al. \cite{kirk83}. The main motivation of
the simulated annealing method comes from the analogy between the
physical annealing of solids and combinatorial optimization
problems. A thorough review of this method can be found in Aarts et al.
\cite{aarts97} or Eglese \cite{eglese90}.

To implement this method, it is necessary to define the problem in
terms of a solution space with a defined neighbourhood and cost
function. The algorithm starts with an initial solution and
moves randomly through the solution space, selecting a neighbour of
the current solution and comparing the difference in cost between
the two solutions. If the new solution provides a better result in
terms of the cost function, we choose it as the new solution.
Otherwise, although it provides a poorer result, there is still a
probability of accepting the new solution. The probability of
accepting a move is typically given by $\exp(-\beta/T)$, where $\beta$
is the cost increment and $T$ is a control parameter
that corresponds to temperature in the analogy with physical annealing.
According to this probability function, small increases in the cost
function are more likely to be accepted than large increases.
Moreover, when $T$ is high, most moves are accepted, whereas when
$T$ approaches zero, the probability of accepting moves is very
small. To avoid local optima, the initial value of $T$ is relatively
high and is gradually decreased when new movements are made.

\subsubsection{Initial solution}
The initial solution is obtained by sorting the groups in ascending
order according to their processing temperature. The samples will be
deposited by columns in the wells according to this group order. In addition, we
must reserve one well for the reagent of the group in all the
plates in which at least one sample of the group is deposited. The reagent
of a group in a plate will be deposited immediately after the last sample
of the group.

Once the samples and reagent of a group have been deposited in the plate,
we can deposit the samples of the next group as close as possible,
under the restriction that samples of both groups can only be in the same strip
if both groups are going to be processed at the same temperature;
otherwise, the samples of the two groups cannot be in the same strip.

If the samples of two consecutive groups must be processed at
different temperatures, once the samples of the first group have been
deposited in the last strip, we deposit the samples of
the next group in the consecutive strip if the difference in the
processing temperature between the two groups does not exceed $5^{0}$ C.
Otherwise, we leave as many empty strips as necessary to satisfy the
constraint on the difference in temperature between consecutive strips.

\begin{algorithm}[ht]
    \DontPrintSemicolon
    \SetNlSty{textbf}{}{.~}
    \SetNlSkip{1mm}
    \SetCommentSty{normal}
    \SetAlgoVlined
    \caption{{\small Initial solution }}
    \label{alg:initialSol}
    \phantom{x}
    \footnotesize
    \KwIn{$samples$, list of samples to be processed; $maxTempDiff$, maximum allowable temperature difference between plate strips}
    \KwOut{$initialSol$, initial solution (list of plates)\;\phantom{x}}
    $plates \leftarrow$~\o\;
    \tcp{Samples are sorted according to their temperature (that is, to the temperature of the group}
    \tcp{each of them belongs to)}
    $sort(samples)$\;
    \For{$i=0$~\emph{\KwTo}~$samples.size$}
    { $posPlate \leftarrow 0$\;
        \tcp{Existing plates are traversed to allocate each sample}
        \While{$posPlate < plates.size$}
        { $currentPlate \leftarrow plates(posPlate)$\;
            $isSampleAssigned \leftarrow \textbf{assignSample}~(samples(i), currentPlate,
            maxTempDiff)$\;
            \If{$!~isSampleAssigned$}
            { $posPlate~+= 1$\; }
            \Else
            { \textbf{break} \; }
        }
        \tcp{Whenever a sample cannot be assigned to existing plates, a new plate is created to contain it}
        \If{$!~isSampleAssigned$}
        { $newPlate \leftarrow new~Plate()$\;
            $\textbf{assignSample}~(samples(i), newPlate, maxTempDiff$)\;
            $addPlate(newPlate, plates)$\;
        }
    }
    $initialSol \leftarrow plates$\;
    \textbf{return}~$initialSol$\;
\end{algorithm}

The pseudocode of the algorithm for obtaining the initial solution is presented as
Algorithm \ref{alg:initialSol}. The main body of the algorithm
describes the initial sorting method. Then, the process of assigning a
sample to existing plates is performed according to the
constraints that are specified above. This process is reflected in the pseudocode in
Algorithm \ref{alg:sampleAssignment}. If a sample cannot be assigned
to an existing plate, a new one is created.


\begin{algorithm}[ht]
    \DontPrintSemicolon
    \SetNlSty{textbf}{}{.~}
    \SetNlSkip{1mm}
    \SetCommentSty{normal}
    \SetAlgoVlined
    \caption{{\small Sample assignment }}
    \label{alg:sampleAssignment}
    \phantom{x}
    \footnotesize
    \KwIn{$sample$, sample to be assigned; $plate$, plate for holding the sample; $maxTempDiff$, maximum allowable temperature difference between plate strips}
    \KwOut{$true$ if $sample$ is finally allocated in $plate$; $false$ otherwise\;\phantom{x}}

    $isSampleAssigned \leftarrow false$\;
    \tcp{First, we attempt to allocate the sample in some of the non-empty strips (that is, strips that already}
    \tcp{hold other samples)}
    $nonEmptyStrips \leftarrow plate.getNonEmptyStrips()$\;
    $posStrip \leftarrow 0$\;
    \While{$(!~isSampleAssigned~~\&\&~~$$posStrip < nonEmptyStrips.size)$}
    { $currentStrip \leftarrow nonEmptyStrips(posStrip)$\;
        \tcp{A sample can be assigned to a non-empty strip if their temperatures match. There must also be}
        \tcp{enough space remaining to hold the sample and the reagent of the group (if needed)}.
        \If{$(\textbf{\upshape checkTemperature}~(sample, currentStrip)~~\&\&~~$$\textbf{\upshape checkSpace}~(sample, currentStrip))$}
        {
            $addSample(sample, currentStrip)$\;
            $isSampleAssigned \leftarrow true$\;
        }
        \Else
        { $posStrip~+= 1$\; }
    }
    \tcp{If none of the non-empty strips can hold the sample, we try to assign it to some of the}
    \tcp{empty strips (if any)}
    \If{$!~isSampleAssigned$}
    {
        $emptyStrips \leftarrow plate.getEmptyStrips()$\;
        $posStrip \leftarrow 0$\;
        \While{$(!~isSampleAssigned~~\&\&~~$$posStrip < emptyStrips.size)$}
        { $currentStrip \leftarrow emptyStrips(posStrip)$\;
            \tcp{To assign a sample to an empty strip, we must guarantee that the maximum allowable }
            \tcp{temperature difference between the plate strips is not exceeded}
            \If{$\textbf{\upshape validateTemperatureDiff}~(sample, currentStrip, plate, maxTempDiff)$}
            {
                $addSample(sample, currentStrip)$\;
                $isSampleAssigned \leftarrow true$\;
            }
            \Else
            { $posStrip~+= 1$\; }
        }
    }
    \textbf{return}~$isSampleAssigned$\;
    \rememberlines
\end{algorithm}

\subsubsection{Movements}

To improve upon the initial solution and obtain the final positions of the
    samples in the plates, the main algorithm explores the search space
    by creating new solutions from a current solution via
    movements. We have devised two movements,
    from which a neighbouring solution is generated:

\begin{enumerate}
    \item[1.] \textit{Grouping movement}. First, a group is randomly selected.
    If there are samples of the same group in different plates, we
    randomly choose two plates where at least one sample of the group
    has been deposited. Then, we try to join all the samples of the group in one of
    the two plates. If the empty wells in the plate are not sufficient
    for this, it is possible to move to the plate samples of other
    groups that have similar processing temperatures. The groups
    that have samples in the other plate are preferred to other groups.
    Moreover, we move the minimum number of plates that are necessary to satisfy the
    feasibility constraints. This movement also allows all the samples of
    one group to be moved from one plate to another plate where there is
    no sample in the group.

    \item[2.] \textit{Strip-exchange movement}. We select two strips in two
    plates and exchange them. Although the temperature constraint
    between consecutive strips should be satisfied for
    the movement to be permitted, the temperatures of the strips can be different.
\end{enumerate}

The movements will be selected according to a fixed parameter that
denotes the probability of each movement. If a movement is not
feasible, a new movement will be made. If no movement is feasible,
the algorithm terminates and the solution that was obtained in the last iteration
is selected as the final solution of the algorithm\footnote{
        This does not limit the search capacity, as we are already
        considering an escape mechanism from local optima based on accepting
        solutions that may be worse than the current one. Instead, this selection avoids
        repeated iterations of the algorithm in a
        rare scenario.}.

\subsubsection{Cost function}

The cost function for comparing the current and new solutions at each
iteration of the algorithm is obtained by applying the lexicographical order
to the following objectives:
\begin{itemize}
    \item First objective: minimize the number of non-empty plates.
    \item Second objective: minimize the number of non-empty wells.
\end{itemize}

We only used two objectives because the definition of the implemented
movements and the second objective indirectly improve the occupancy
rate of the plates. Furthermore, we could have considered a single objective, as in the ILP model.
However, the use of a single aggregated objective often results in
the heuristic algorithm finding local minima and in a high
degree of variability in the results. Thus, we have ruled out this
strategy.

The pseudocode of the algorithm for obtaining the final solution is presented as
Algorithm \ref{alg:simannel}.


\begin{algorithm}[htp]
    \DontPrintSemicolon
    \SetNlSty{textbf}{}{.~}
    \SetNlSkip{1mm}
    \SetCommentSty{normal}
    \SetAlgoVlined
    \caption{{\small Simulated annealing}}
    \label{alg:simannel}
    \phantom{x}
    \footnotesize
    \KwIn{$initialSol$, initial solution; $maxIter$, maximum number of algorithm iterations;
        $Tmax$, initial temperature; $Tmin$, minimum temperature; $\alpha$, cooling parameter;
        $SEProb$, probability of applying a $Strip-exchange$ movement (thus, a $Grouping$
        movement will be applied with probability $1-SEProb$)}
    \KwOut{$bestSol$, best global solution that has been found\;\phantom{x}}

    $bestSol \leftarrow initialSol$\;
    $currentSol \leftarrow initialSol$\;
    \For{$i=0$~\emph{\KwTo}~$maxIter$}
    { $T \leftarrow Tmax$\;
        \While{$T \geq~Tmin$}
        { \tcp{A new solution is created from the current one by applying a movement. If}
            \tcp{no movement can be applied, the algorithm terminates}
            $newSol \leftarrow \textbf{applyMovement}~(currentSol, SEProb)$\;
            \If{$newSol =$ \o}
            { \textbf{return}~$bestSol$\; }
            \Else
            { \tcp{We must always keep updated the best global solution that has been found so far}
                $updateBestSol~(newSol, bestSol)$\;
                \tcp{If $newSol$ improves $currentSol$ (that is, if it uses fewer plates or}
                \tcp{if it minimizes the number of non-empty wells), it will replace the current solution}
                \tcp{ for the next iteration}
                \Case{$\textbf{\upshape numPlates}~(newSol) < \textbf{\upshape numPlates}~(currentSol)$}
                { $currentSol \leftarrow newSol$\;}

                \Case{$\textbf{\upshape numPlates}~(newSol) = \textbf{\upshape numPlates}~(currentSol)$}
                {
                    \If{$\textbf{\upshape nonEmptyWells}~(newSol) \leq \textbf{\upshape nonEmptyWells}~(currentSol)$}
                    { $currentSol \leftarrow newSol$\;}
                    \Else
                    {
                        $threshold \leftarrow e^{\frac{nonEmptyWellsDiff(newSol,
                                currentSol)}{T}}$\;
                        $\textbf{acceptWorseSolution}~(generateRandom(), threshold, newSol, currentSol)$\;
                    }
                }
                \tcp{Even when $newSol$ is worse than $currentSol$, it still may replace the current solution}
                \Other
                {
                    $threshold \leftarrow e^{\frac{numPlatesDiff(newSol,currentSol)}{T}}$\;
                    $\textbf{acceptWorseSolution}~(generateRandom(), threshold, newSol, currentSol)$\;
                }
            }
            $T \leftarrow \alpha * T$\;
        }
    }
    \textbf{return}~$bestSol$\;

\end{algorithm}

\subsubsection{Example}

\begin{figure}[htp]
    \centering
    \includegraphics[scale=0.55]{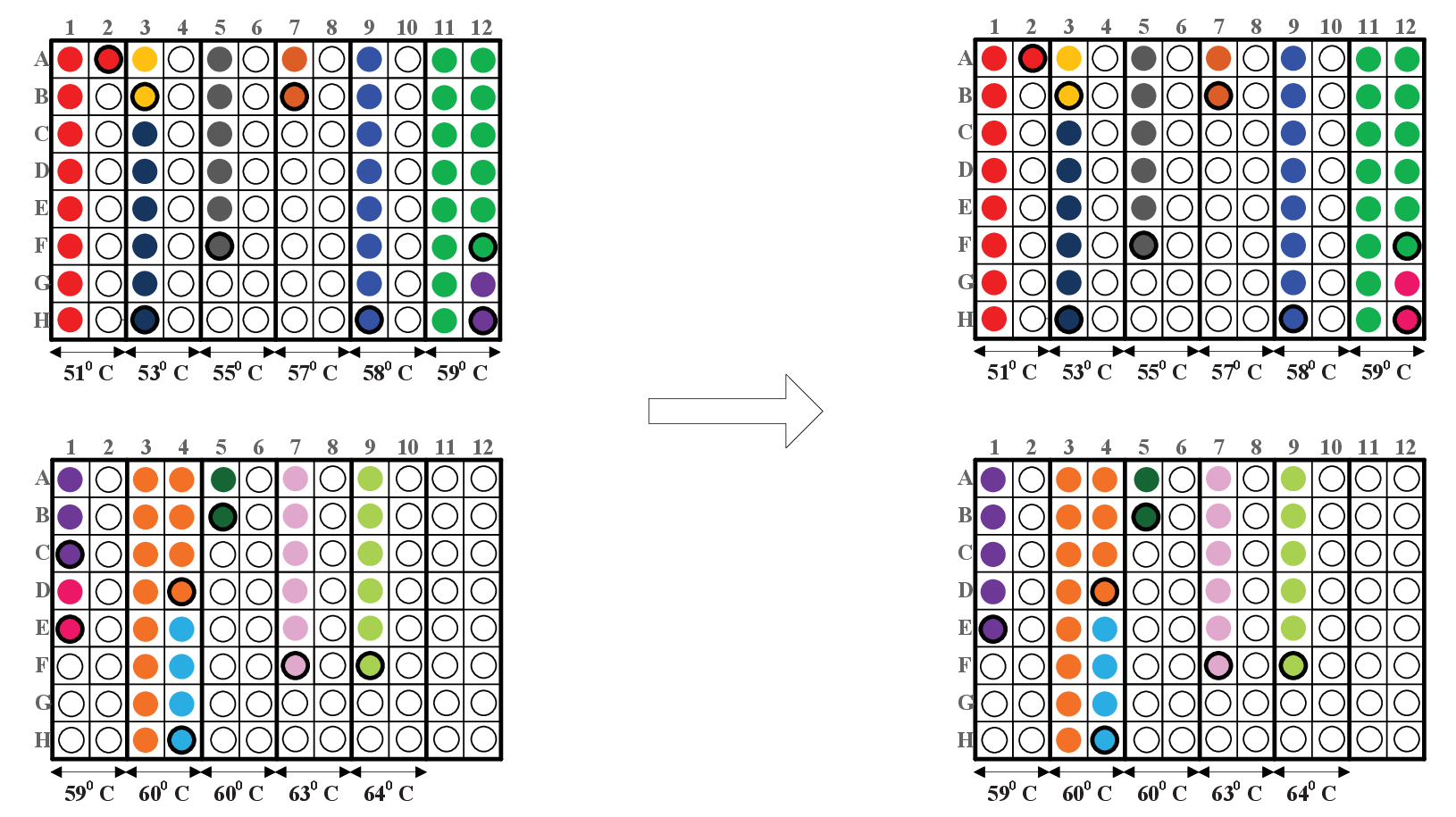}
    \caption{Grouping movement} \label{fig:first_movement}
\end{figure}

To demonstrate how the algorithm operates, we consider the following example: We must distribute 69 samples that are organized in 14 groups (and the corresponding reagents) in the plates. The initial solution of the algorithm is represented in the first part of Figure \ref{fig:first_movement}. The groups are ordered according to their temperatures in increasing order. According to this order, we fix the temperatures of the strips. To organize the samples under this criterion, we need two plates and the number of occupied cells is 84 (69 samples and 15 reagents). There are several strips that have the same temperature since in these cases, one strip is not sufficient to hold all the samples that are processed at the corresponding temperature. Moreover, the samples of the \textquotedblleft violet\textquotedblright\textsf{ }group are distributed in both plates. Then, the grouping movement could consist of joining all the samples of the group in the first plate and moving the \textquotedblleft pink\textquotedblright\textsf{ }group to the first plate.

Both solutions (the initial solution and the solution obtained after 1 iteration of applying the grouping movement of the algorithm) require two plates. However, whereas the number of occupied cells in the initial solution is 84, this number is reduced to 83 after 1 iteration since the number of reagents has been reduced from 15 to 14. Thus, the solution has been improved after 1 iteration.

\begin{figure}[htp]
    \centering
    \includegraphics[scale=0.55]{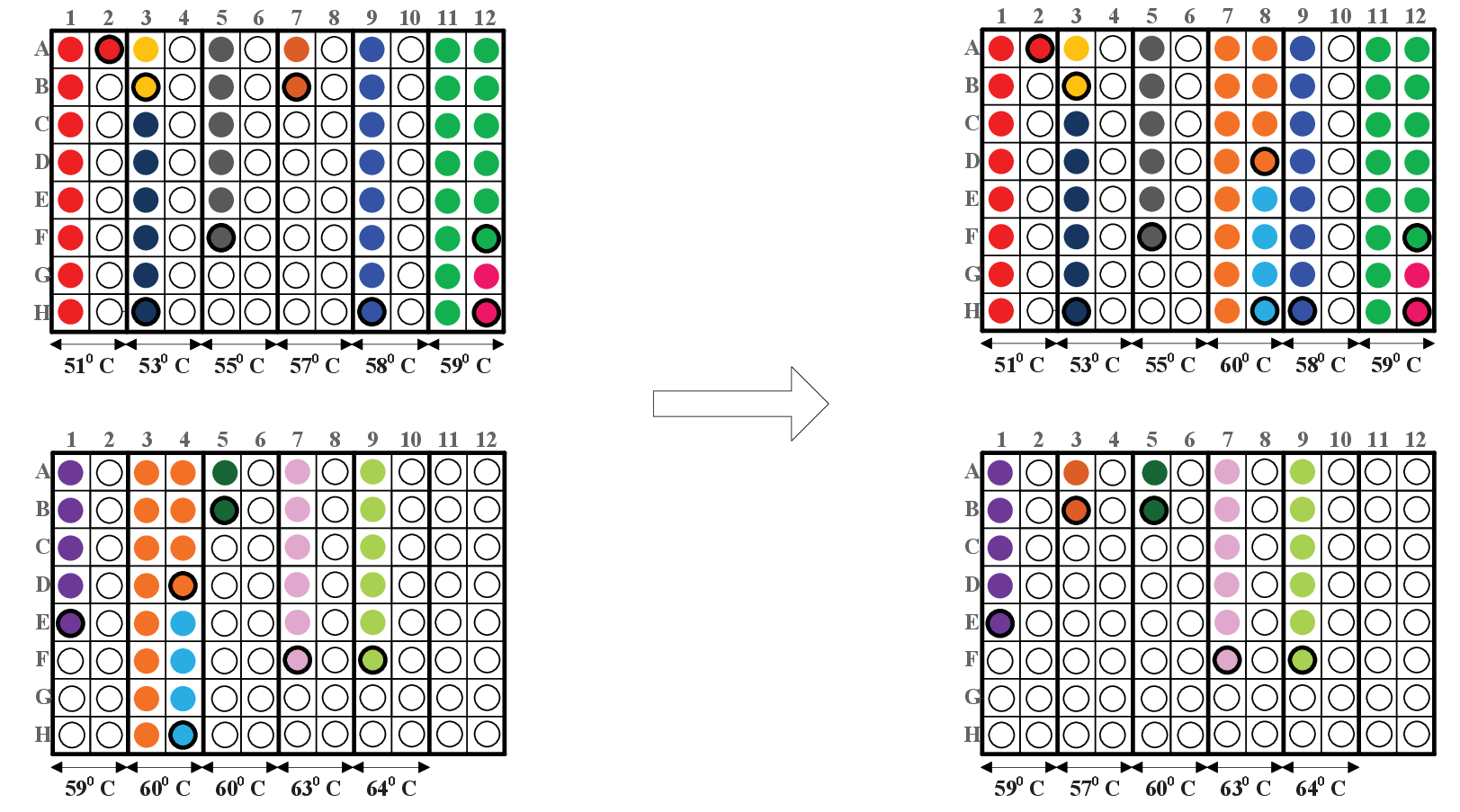}
    \caption{Strip-exchange movement} \label{fig:second_movement}
\end{figure}

Suppose we accept this new solution and perform another iteration. Now, we could apply the strip-exchange movement on the fourth strip of the first plate and the second strip of the second plate. After this movement, the number of occupied cells is again 83; however, the solution is improved because the occupancy rates have changed from (51.04,35.42) to (65.63,20.83). This movement is illustrated in Figure \ref{fig:second_movement}.

\section{Experimental results}

\subsection{Data}
The solution methods have been tested with real data. We have considered 36 files, which are divided into two sets:
\begin{itemize}
    \item The first set contains 30 files (Id 1-30), each of which contains information on one working session in the laboratory.
    \item The second set contains 6 files (Id 31-36). Each one of these files corresponds to 2 or 3 plates that have been
    randomly chosen from one working session.
\end{itemize}
Table \ref{tabla1} summarizes the main characteristics of the files (numbers of samples, groups and group temperatures). In both sets, the files have been ordered according to the number of samples. The number of temperatures remains stable in all scenarios. However, the number of groups and, especially, the number of samples vary considerably. Hence, the characteristics of each working session are highly particular.

\begin{table}[htp]
    \centering
        \begin{tabular}{lllllllllllll}
        \hline
        Id & 1 & 2 & 3 & 4 & 5 & 6 & 7 & 8 & 9 & 10 & 11 & 12\\
        \hline
        Samples & 174 & 193 & 233 & 285 & 290 & 315 & 358 & 368 & 432 & 434 & 501 & 551\\
        \hline
        Groups & 80 & 68 & 31 & 147 & 87 & 99 & 35 & 27 & 32 & 44 & 82 & 37\\
        \hline
        Temperatures & 13 & 14 & 12 & 18 & 10 & 16 & 10 & 11 & 14 & 13 & 15 & 15\\
        \hline\hline
        Id & 13 & 14 & 15 & 16 & 17 & 18 & 19 & 20 & 21 & 22 & 23 & 24\\
        \hline
        Samples & 612 & 647 & 747 & 797 &  876 & 918 & 963 & 1128 & 1270 & 1309 & 1398 & 1473\\
        \hline
        Groups & 107 & 27 & 37 & 27 & 120 & 184 & 37 & 192 & 167 & 201 & 200 & 197\\
        \hline
        Temperatures & 17 & 12 & 15 & 11 & 13 & 17 & 15 & 16 & 17 & 17 & 18 & 16\\
        \hline\hline
        Id & 25 & 26 & 27 & 28 & 29 & 30 & 31 & 32 & 33 & 34 & 35 & 36\\
        \hline
        Samples & 1944 & 2071 & 2248 & 2496 & 2703 & 3783 & 40 & 44 & 50 & 65 & 77 & 84\\
        \hline
        Groups & 151 & 162 & 165 & 179 & 200 & 171 & 24 & 31 & 45 & 34 & 43 & 68\\
        \hline
        Temperatures & 15 & 15 & 17 & 17 & 17 & 17 & 11 & 11 & 14 & 15 & 13 & 16\\
        \hline
    \end{tabular}
\caption{Description of the files that were obtained from real data}
\label{tabla1}
\end{table}

\subsection{Comparison between LabWare, the ILP model and the heuristic algorithm}

This section is devoted to comparing the results that were obtained in the laboratory with LabWare software\footnote{Since LabWare is a comercial software, we do not know exactly how the optimization tool of this software works.} and the results that were obtained via our solution methods. Our solution methods have been evaluated using a computer with an Intel(R) Core(TM) i7-3770 CPU of 3.40 GHz and 16 GB of RAM. Moreover, the ILP problem was solved using Gurobi Optimizer (version 6.0.5), whereas the heuristic algorithm was implemented in Java.

Each working session in the laboratory involves many samples and groups. Thus, the ILP problem for Files 1-30, which are related to real working sessions, will be large and cannot be solved in a reasonable amount of time.
To demonstrate this, Table \ref{tabla2} lists the results that were obtained with ten examples from the first set. The ILP problem solutions were obtained after 60 hours.
According to Table \ref{tabla2}, after this time, we cannot guarantee the optimality of the solutions. The table also indicates the best solution that was obtained and the size\footnote{When we refer to the size of an ILP problem, we first indicate the number of constraints, followed by the number of variables.} of the ILP problem, the gap of the solution\footnote{The gap is computed with the following formula:  $100*\frac{|objective.lower.bound-incumbent.objective|}{|incumbent.objective|}$.}  and the minimum time needed to obtain the solution.
The results demonstrate the necessity of developing a tool to identify satisfactory solutions in a reasonable computational time.

\begin{table}[htp]
    \centering
    \begin{tabular}{llllll}
        \hline
        \multirow{2}{*}{Id} & \multicolumn{5}{c}{ILP problem solution} \\
        \cline{2-6}
        & Plates & Occupancy rates & Size & Gap (\%) & Minimal time\\
        \hline
        1 & 4 & 100/98.96/44.79/22.92 & 4664$\times$4176 & 2.17 & 2.07 h\\
        \hline
        2 & 5 & 100/97.92/55.21/20.83/2.08 & 5048$\times$4530 & 3.14 & 0.21 h\\
        \hline
        3 & 4 & 100/94.79/64.58/26.04 & 2043$\times$1800 & 4.17 & 54.87 h\\
        \hline
        4 & 7 & 98.96/97.92/96.88/95.83/39.58/17.71/6.25 & 14476$\times$13146 & 5.43 & 59.61 h\\
        \hline
        5 & 5 & 100/100/100/72.92/22.92 & 6182$\times$5550 & 1.18 & 12.89 h\\
        \hline
        6 & 6 & 100/100/96.87/84.37/46.87/6.25 & 7989$\times$7452 & 1.02 & 3.46 h\\
        \hline
        7 & 5 & 100/100/100/80.20/36.46 & 2750$\times$2430 & 0.99 & 58.21 h\\
        \hline
        8 & 5 & 100/100/98.96/86.46/31.25 & 2252$\times$1980 & 1.90 & 53.8 h\\
        \hline
        9 & 7 & 100/100/100/98.96/65.63/23.96/2.08 & 3364$\times$3318 & 1.90 & 41.05 h\\
        \hline
        10 & 7 & 100/100/98.96/98.96/75.00/28.13/4.17 & 4790$\times$4284 & 1.79 & 45.92 h\\
        \hline
    \end{tabular}
    \caption{Solutions for Files 1-10 that were obtained by solving the ILP problem}
    \label{tabla2}
\end{table}

However, it is possible to find the optimal solutions in few seconds for problems that are associated with a small number of plates. For
this reason, we have considered 6 small examples with 2 or 3 plates that are
configured in the laboratory with LabWare. Despite not being representatives of a typical working session, these examples will be useful for comparing optimal solutions with the solutions that are provided by the heuristic algorithm. Thus, in Table \ref{tabla3} we can compare the occupancy rates of the three methods: LabWare, the ILP model and the heuristic algorithm. The table also lists the computational times for the ILP model and the heuristic algorithm.

\begin{table}[htp]
    \centering
    \begin{tabular}{llllllll}
        \hline
        Id & \multicolumn{3}{c}{Occupancy rates} & & \multicolumn{3}{c}{Computational times}\\
        \cline{2-4}\cline{6-7}
        & LabWare & ILP model & Heuristic & & ILP model & Heuristic\\
        \hline
        31 & 37.50/29.17 & 50.00/16.67  & 50.00/16.67 & & 0.01 s & 0.006 s \\
        \hline
        32 & 29.17/28.13/20.83 & 57.29/20.83 & 57.29/20.83 & & 0.25 s & 0.07 s\\
        \hline
        33 & 54.17/37.50/7.29 & 71.88/20.83/6.25 & 71.88/20.83/6.25 & & 0.88 s  & 0.065 s \\
        \hline
        34 & 41.67/39.58/21.88 & 73.96/20.83/8.33  & 73.96/20.83/8.33 & & 4.04 s & 0.26 s \\
        \hline
        35 & 47.92/47.92/30.21 & 86.46/32.29/6.25 & 86.46/32.29/6.25 & & 4.76 s & 0.55 s  \\
        \hline
        36 & 67.71/50.00/40.63 & 91.67/45.83/20.83 & 91.67/45.83/20.83 &  & 4.76 s & 0.62 s \\
        \hline
    \end{tabular}
    \caption{Examples with 2 or 3 plates}
    \label{tabla3}
\end{table}

We have obtained the
optimal solutions of the ILP problem in Files 31-36. However, the solution  provided by LabWare never coincides with the optimal solution. In the second example, the LabWare solution
requires an additional plate and in the other cases, the distribution of the samples in the plates is worse. However, after determining the best values for the input parameters in the algorithm (a study of the parameters of the algorithm will be performed next), we have verified that for these $2$ and $3$ plate files, the heuristic algorithm reaches the optimal value provided by the ILP model in less time. Since Files 31-36 have been randomly selected from the same working session, these results enable us to guarantee the satisfactory performance of the heuristic algorithm.

\subsection{Study of the parameters in the heuristic algorithm}
In this section, we focus on obtaining the appropriate parameters
for Algorithm 3 according to our experimental environment.

The algorithm depends on several input parameters, as discussed above.  To ensure its proper performance in our scenario, it is necessary to adjust these parameters according to our data. Next, we justify the values of these input parameters:
\begin{itemize}
    \item Initial temperature. The initial temperature was set to 100. According to Ben-Ameur \cite{benAmeur2004}, this value ensures a low probability of acceptance of poor solutions with respect to the first objective and higher acceptance ratio with respect to the second objective. Thus, when the number of plates is minimized, there is greater flexibility for filling them in the best possible way.
    \item Minimum temperature. The final temperature at each simulated annealing iteration was set to $1E-10$. This low value permits an exhaustive search at each iteration.

     \item Cooling parameter ($\alpha$). Our annealing procedure initially considers a high temperature. Then, the temperature is lowered incrementally by a constant factor. It is important to consider sufficiently many steps at each temperature to keep the system close to equilibrium until the system approaches the minimum temperature. We have considered $\alpha= 0.9$.

    \item Maximum number of iterations. Once the above parameters were set, we performed an exhaustive study of the execution times by varying the number of iterations. According to the results of this study, a value of 1000 iterations has been chosen to ensure a reasonable execution time in all scenarios. In addition, the objective function values of the problem do not improve after 1000 iterations.

    \item Probability of the $Strip$-$exchange$ movement, which is the parameter that most influences the result of the algorithm. We have studied the results that were obtained for all files by varying the probability of choosing this movement from $0$ to $1$ in increments of $0.1$. According to this empirical study, this movement is the more effective of the two implemented movements as it obtains only the best solutions if a high probability has been assigned to it.

     For simplicity, Table \ref{tab-exmov} lists the results for probability values of $0.8$, $0.9$ and $1$. To summarize the results for each instance, Table \ref{tab-exmov} lists the minimum number of plates that are required for placing the samples, the total number of full plates and the fill rate of the first plate that is not filled. The $Grouping$ movement plays no role in small instances, as poorer results are always obtained when it is involved. However, for larger files, it is important to apply it with low probability to obtain the best results. For instance, in File $27$, we can obtain $24$ full plates out of a total of $27$ by assigning probability $0.2$ to the $Grouping$ movement and probability $0.8$ to the $Strip$-$exchange$ movement.

      \begin{figure}[H]
        \centering
        \subfigure{\includegraphics[scale=0.3]{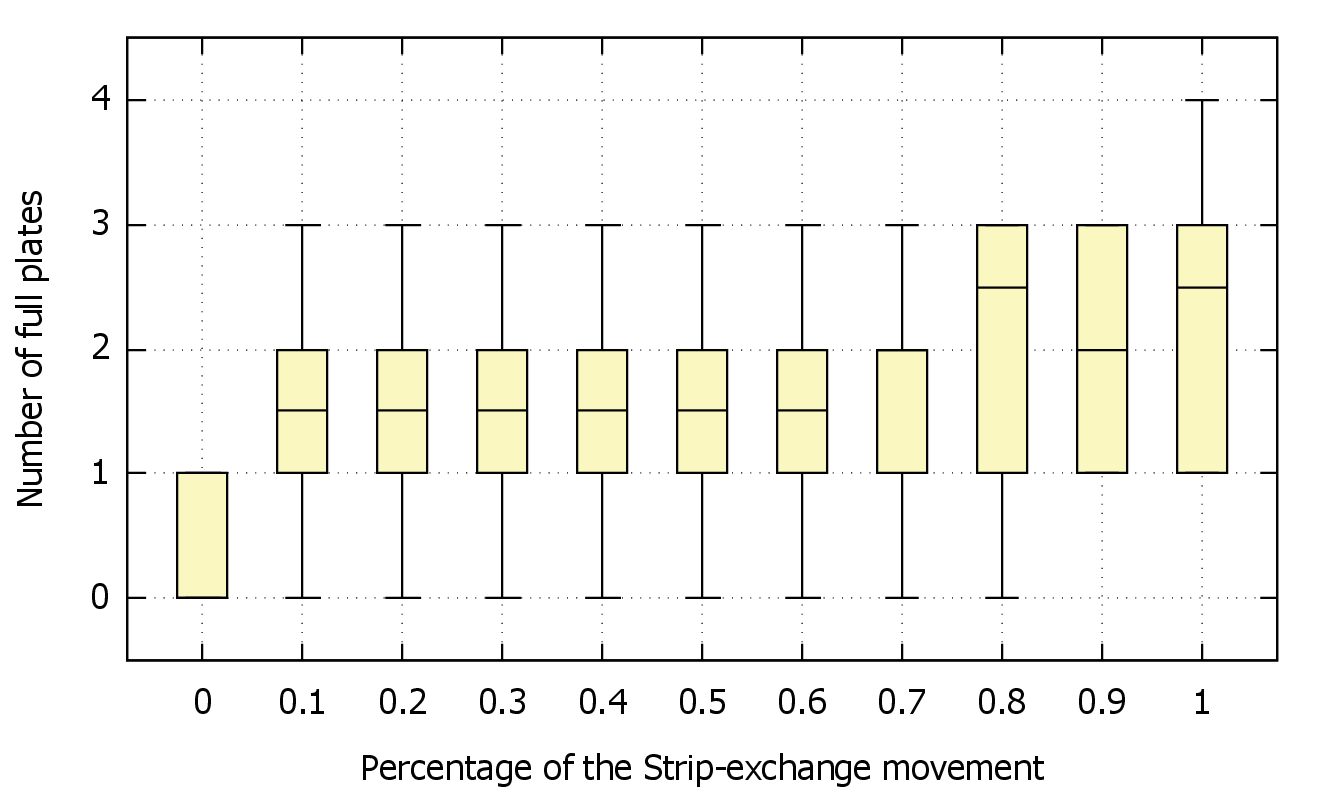}}
        \subfigure{\includegraphics[scale=0.3]{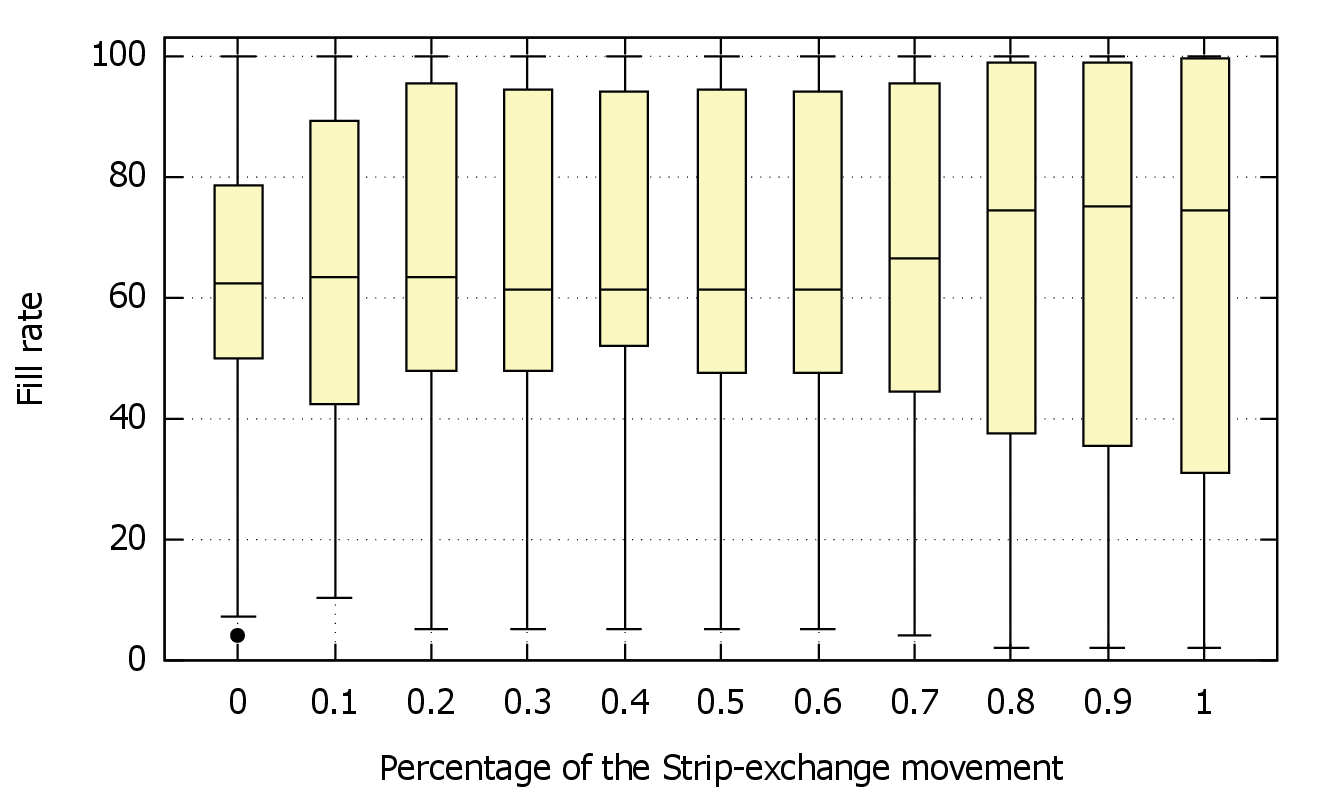}}
        \caption{Study of the Strip-exchange movement for small instances (Files 1-10)}
        \label{peq}
     \end{figure}

     \begin{figure}[H]
        \centering
        \subfigure{\includegraphics[scale=0.3]{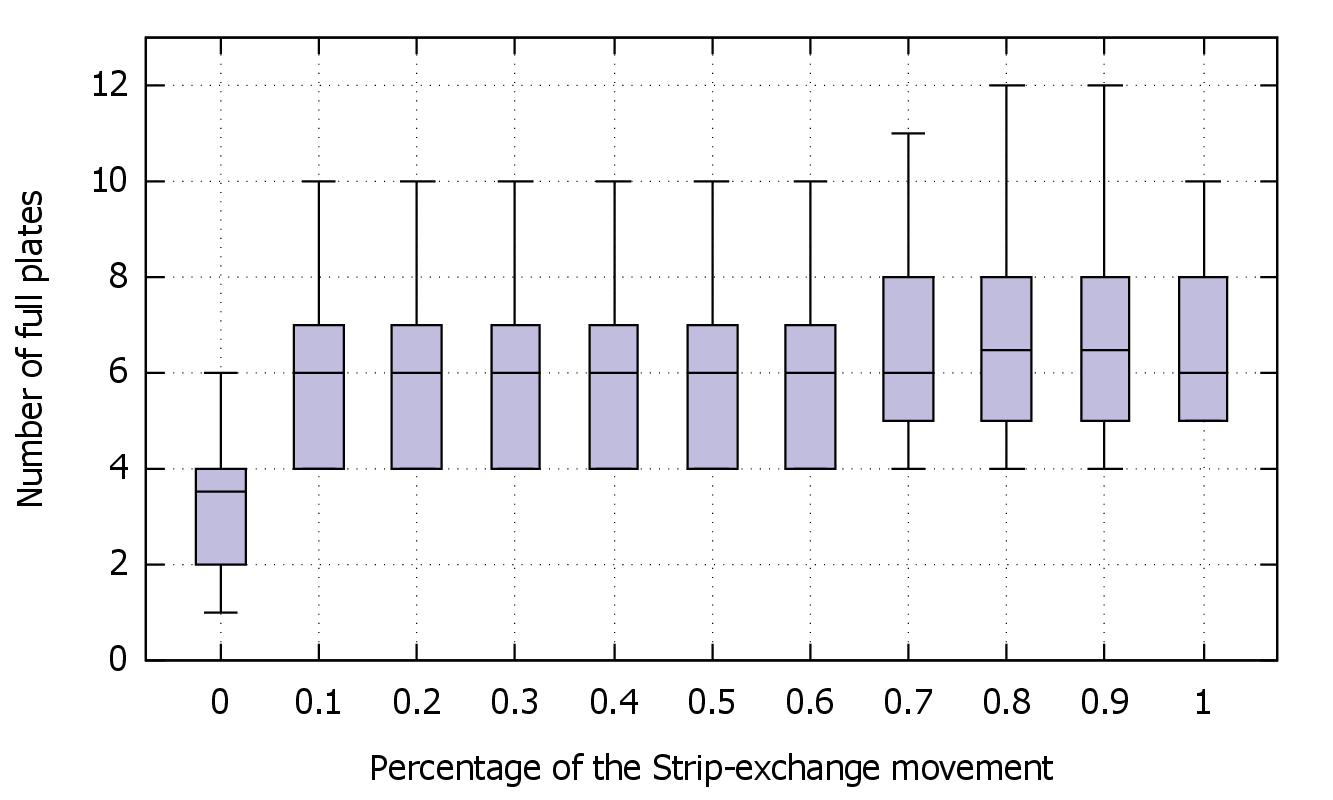}}
        \subfigure{\includegraphics[scale=0.3]{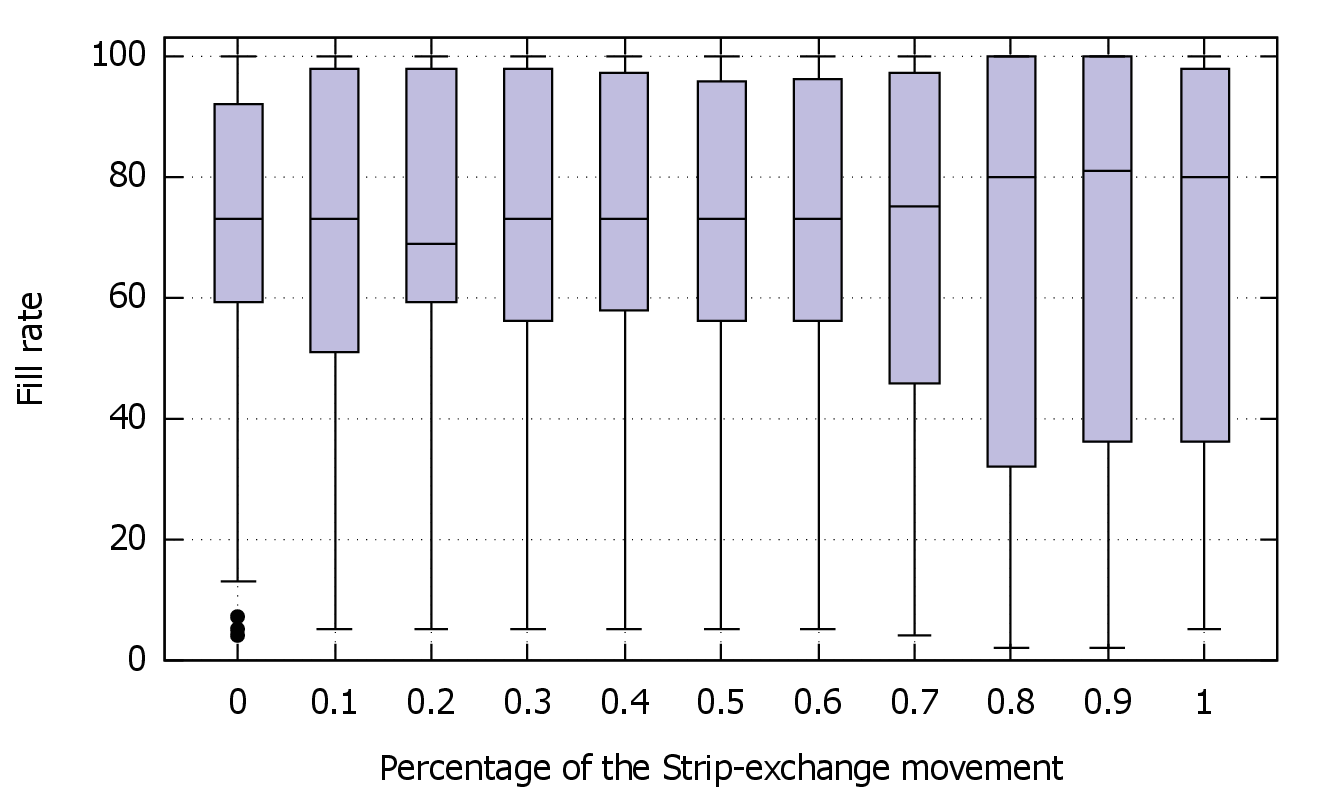}}
        \caption{Study of the Strip-exchange movement for medium instances (Files 11-20)}
        \label{med}
     \end{figure}

     \begin{figure}[H]
        \centering
        \subfigure{\includegraphics[scale=0.3]{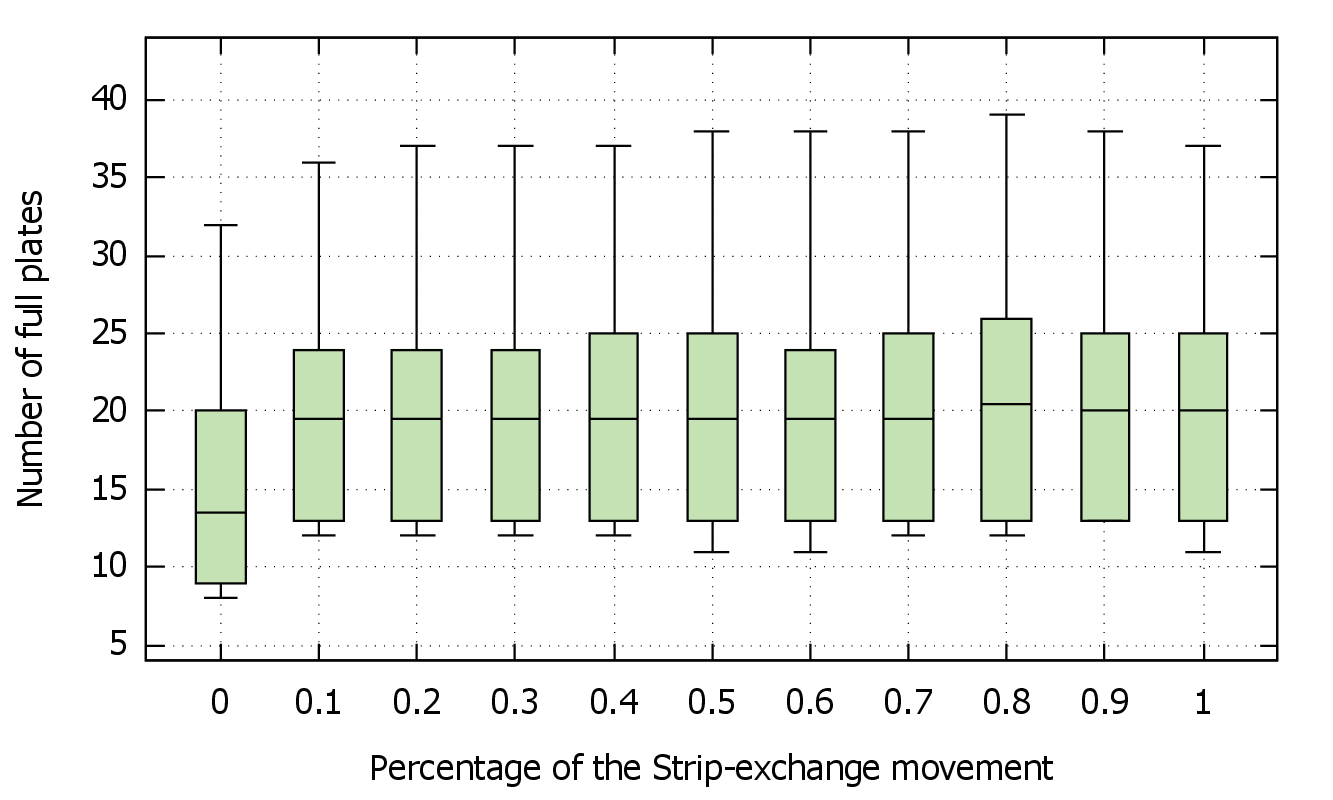}}
        \subfigure{\includegraphics[scale=0.3]{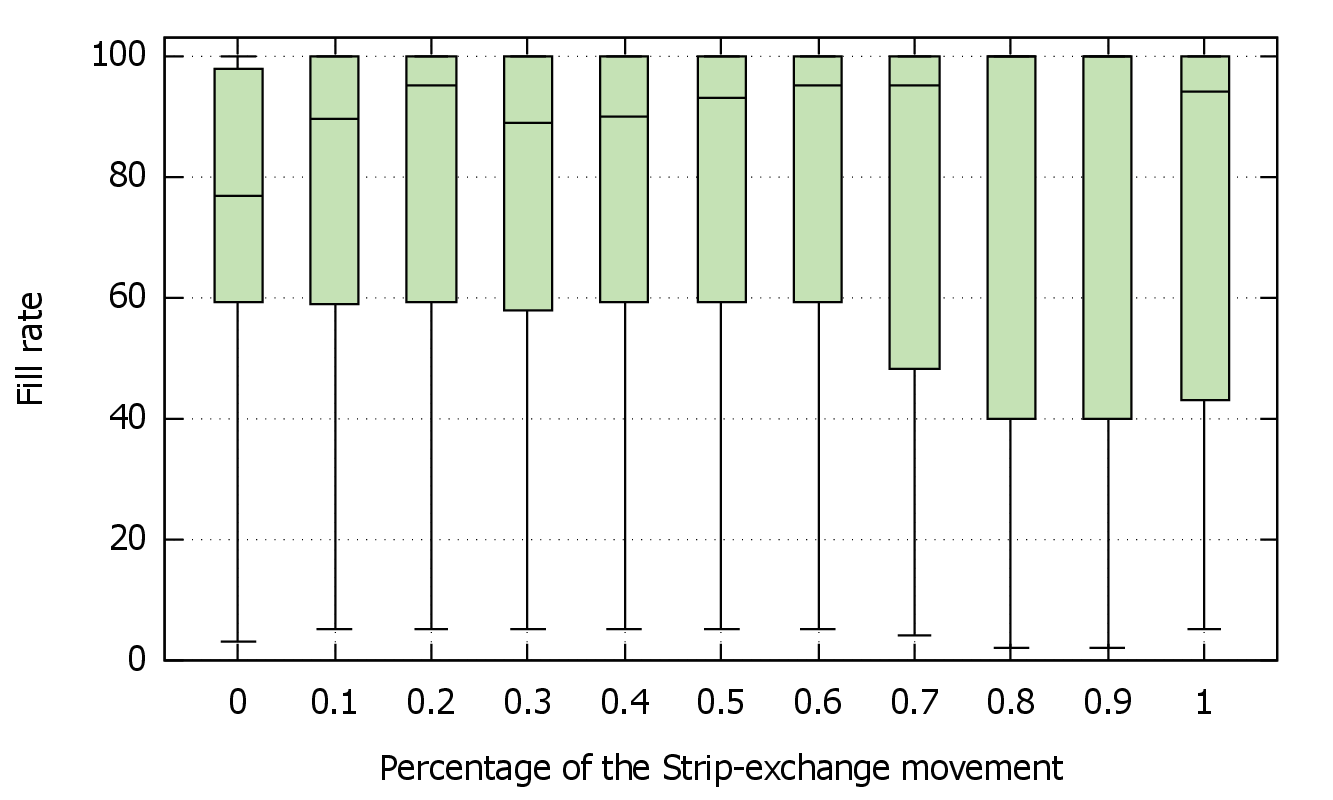}}
        \caption{Study of the Strip-exchange movement for large instances (Files 21-30)}
        \label{big}
     \end{figure}

      Figures \ref{peq}, \ref{med} and \ref{big} show the fill rates that were obtained for various probabilities of the $Strip$-$exchange$ movement in small, medium and large instances, respectively. Figures \ref{peq}, \ref{med} and \ref{big} support the following observation from Table \ref{tab-exmov}: the probabilities of 0.8 and 0.9 for the $Strip$-$exchange$ movement yield the best filling rates for medium and large working sessions.

    \begin{table}[htbp]
        \centering

        \begin{tabular}{llllllll}
            \hline
            \multicolumn{1}{l}{\multirow{2}[4]{*}{Id}} & \multicolumn{1}{l}{\multirow{2}[4]{*}{Plates}} & \multicolumn{3}{c}{Full plates} & \multicolumn{3}{c}{Percentage} \\
            \cline{3-8}          &       & \multicolumn{1}{l}{ 0.8} & \multicolumn{1}{l}{  0.9} & \multicolumn{1}{l}{ 1} &  0.8 &  0.9 &  1 \\
            \hline
            1     & 4     & 1     & 1     & 1     & 96.88 & 97.92 & 98.96 \\
            \hline
            2     & 5     & 1     & 1     & 1     & 91.67 & 86.46 & 95.83 \\
            \hline
            3     & 4     & 0     & 1     & 1     & 97.92 & 80.21 & 80.21 \\
            \hline
            4     & 7     & 1     & 1     & 2     & 97.92 & 97.92 & 97.92 \\
            \hline
            5     & 5     & 2     & 2     & 3     & 91.67 & 91.67 & 75 \\
            \hline
            6     & 6     & 3     & 2     & 2     & 85.42 & 93.75 & 98.96 \\
            \hline
            7     & 5     & 3     & 3     & 3     & 80.21 & 80.21 & 81.25 \\
            \hline
            8     & 5     & 3     & 3     & 3     & 85.42 & 84.38 & 86.46 \\
            \hline
            9     & 7     & 3     & 3     & 4     & 94.79 & 96.88 & 39.58 \\
            \hline
            10    & 7     & 3     & 3     & 4     & 95.83 & 95.83 & 63.54 \\
            \hline
            11    & 8     & 5     & 5     & 5     & 54.17 & 79.17 & 76.04 \\
            \hline
            12    & 8     & 5     & 5     & 5     & 84.38 & 79.17 & 73.96 \\
            \hline
            13    & 9     & 4     & 4     & 5     & 98.96 & 98.96 & 91.67 \\
            \hline
            14    & 9     & 6     & 6     & 6     & 83.33 & 85.42 & 78.12 \\
            \hline
            15    & 10    & 6     & 6     & 6     & 98.96 & 98.96 & 91.67 \\
            \hline
            16    & 10    & 7     & 7     & 6     & 94.79 & 89.58 & 97.92 \\
            \hline
            17    & 12    & 8     & 8     & 8     & 98.96 & 98.96 & 97.92 \\
            \hline
            18    & 13    & 8     & 8     & 8     & 95.83 & 98.96 & 95.83 \\
            \hline
            19    & 12    & 9     & 9     & 8     & 89.58 & 94.79 & 93.75 \\
            \hline
            20    & 15    & 12    & 12    & 10    & 86.46 & 96.88 & 96.88 \\
            \hline
            21    & 17    & 13    & 13    & 11    & 85.42 & 93.75 & 97.92 \\
            \hline
            22    & 18    & 12    & 13    & 12    & 98.96 & 98.96 & 98.96 \\
            \hline
            23    & 19    & 13    & 13    & 13    & 95.83 & 97.92 & 88.54 \\
            \hline
            24    & 19    & 14    & 15    & 14    & 97.92 & 82.29 & 97.92 \\
            \hline
            25    & 23    & 20    & 19    & 19    & 93.75 & 98.96 & 83.33 \\
            \hline
            26    & 25    & 21    & 21    & 21    & 98.96 & 98.96 & 93.75 \\
            \hline
            27    & 27    & 24    & 23    & 21    & 76.04 & 97.92 & 85.42 \\
            \hline
            28    & 30    & 26    & 25    & 25    & 98.96 & 94.79 & 93.75 \\
            \hline
            29    & 32    & 29    & 28    & 28    & 87.5  & 96.88 & 86.46 \\
            \hline
            30    & 44    & 39    & 38    & 37    & 98.96 & 86.46 & 96.98 \\
            \hline
        \end{tabular}%
        \caption{Results of applying the $Strip$-$exchange$ movement with probabilities $0.8$, $0.9$ and $1$}
        \label{tab-exmov}%
    \end{table}

\end{itemize}

\subsection{Final results}

Once the parameters of the algorithm have been chosen, we will compare the results that were obtained by applying the heuristic algorithm with those that were obtained in the laboratory using the LabWare software for Files 1-30.

\begin{figure}[h]
    \centering
    \includegraphics[width=10cm]{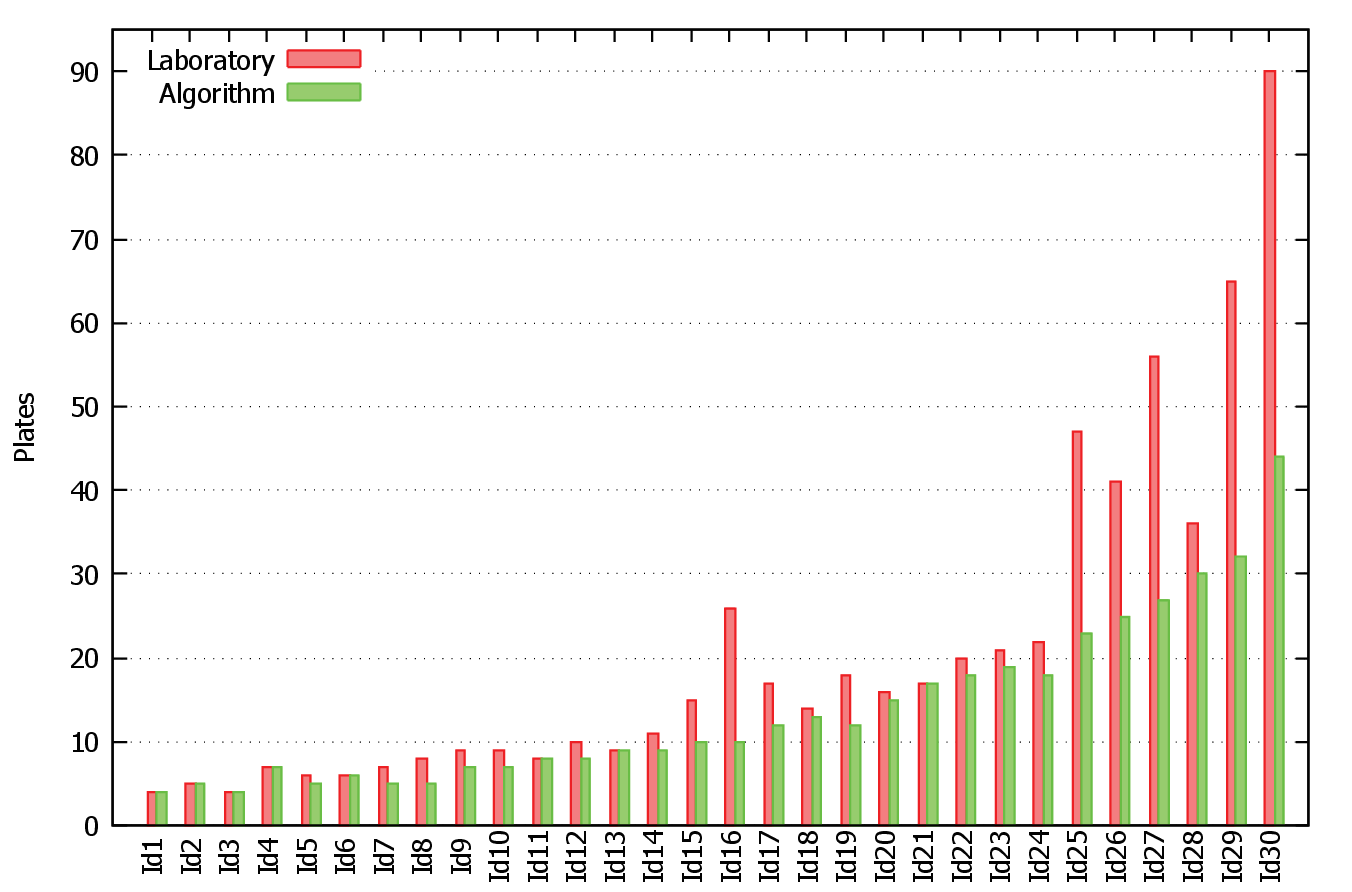}
    \caption{Comparison between the total number of plates that were obtained via the LabWare software and the heuristic algorithm} \label{plates}
\end{figure}

The first and most important objective is the reduction of the number of plates necessary to process the samples. Figure \ref{plates} shows the number of plates that are saved by the heuristic algorithm compared to the solution that was provided by the LabWare software. In view of  Figure \ref{plates}, it is evident that the heuristic algorithm is able to provide better solutions in most of the files. The savings are more significant in large instances.

\begin{figure}[h]
    \centering
    \includegraphics[width=10cm]{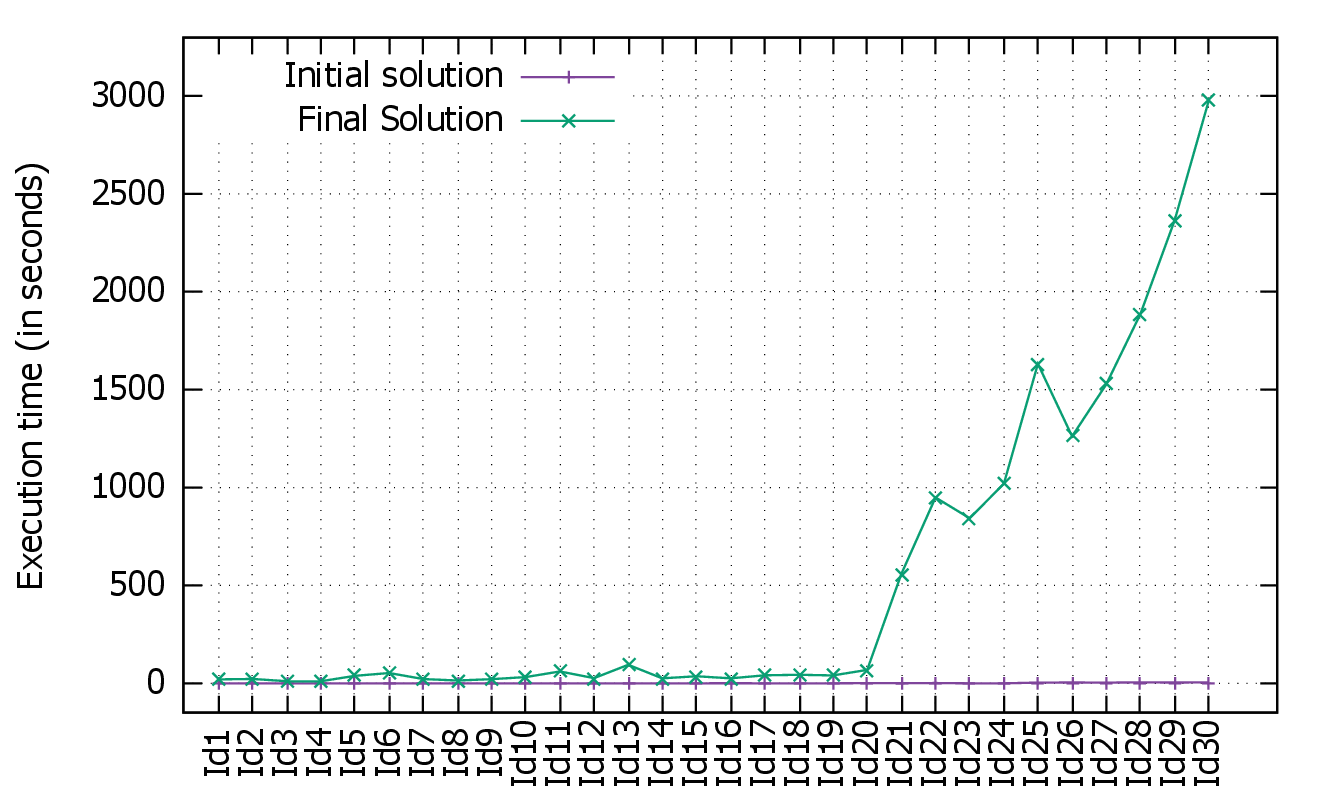}
    \caption{Comparison of the times in which the initial and final solutions of the algorithm are obtained} \label{time}
\end{figure}

In addition, the heuristic algorithm achieves the solutions in reasonable times for the laboratory, as shown in Figure \ref{time}. It is also provided a comparison of the times necessary for the heuristic algorithm to obtain the initial and final solutions. The initial solution is calculated in few seconds for all working sessions, whereas the computational time of the final solution depends to a large extent on the number of samples of each file.

\begin{figure}[h]
    \centering
    \subfigure{\includegraphics[scale=0.31]{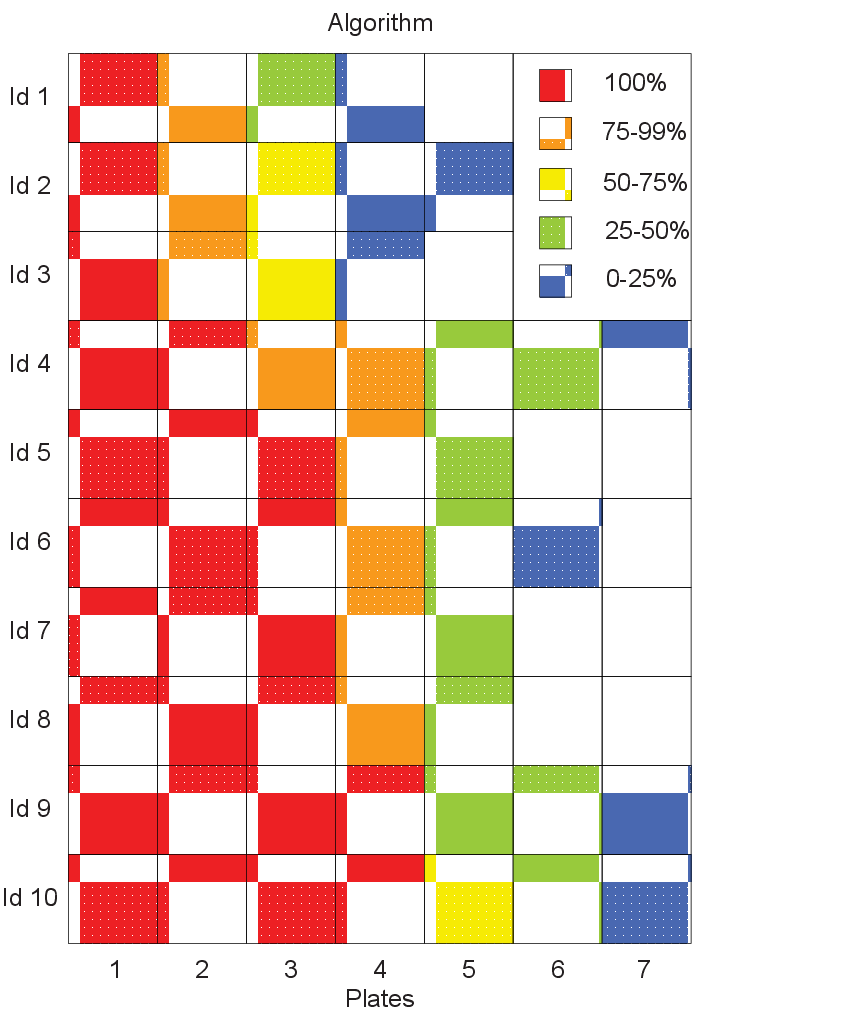}}
    \hspace*{2cm}
    \subfigure{\includegraphics[scale=0.31]{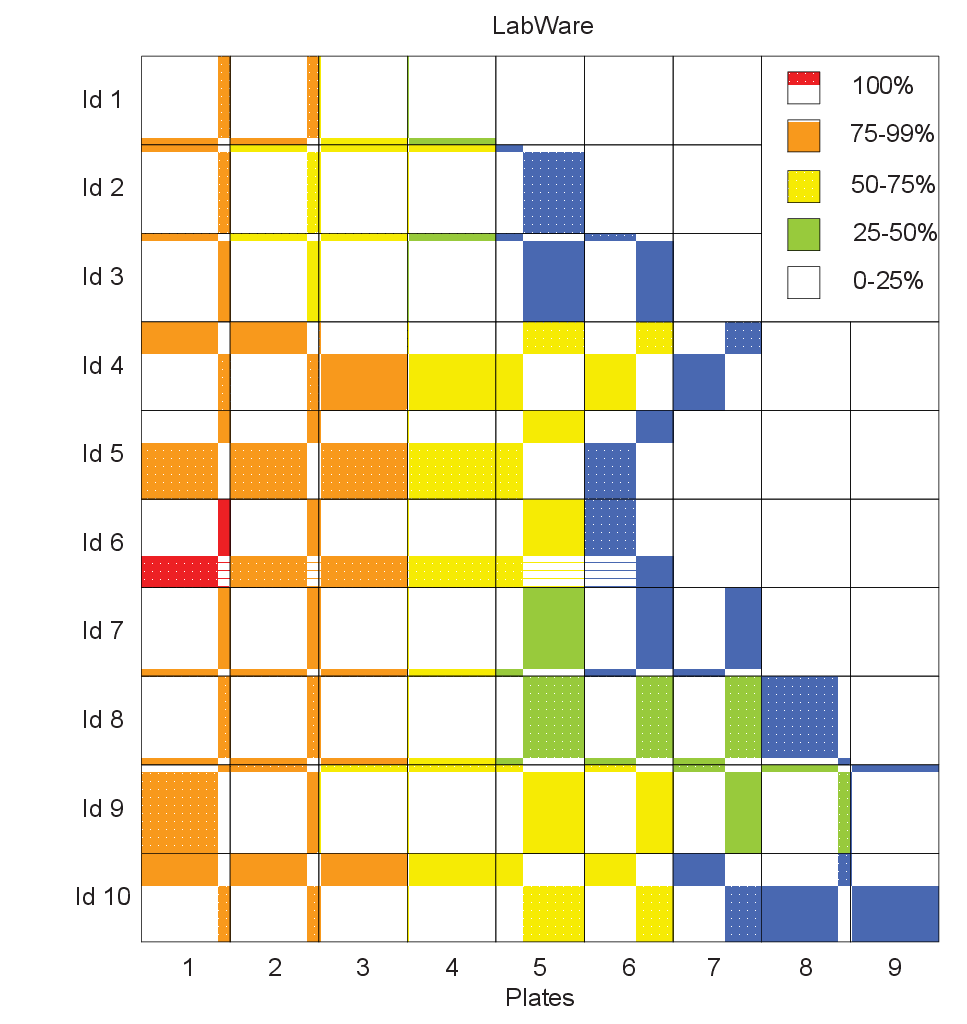}}
    \caption{Distribution of the plate occupancy obtained with the algorithm and LabWare in small instances} \label{mapacolor1}
\end{figure}

\begin{figure}[h]
    \centering
\includegraphics[scale=0.43]{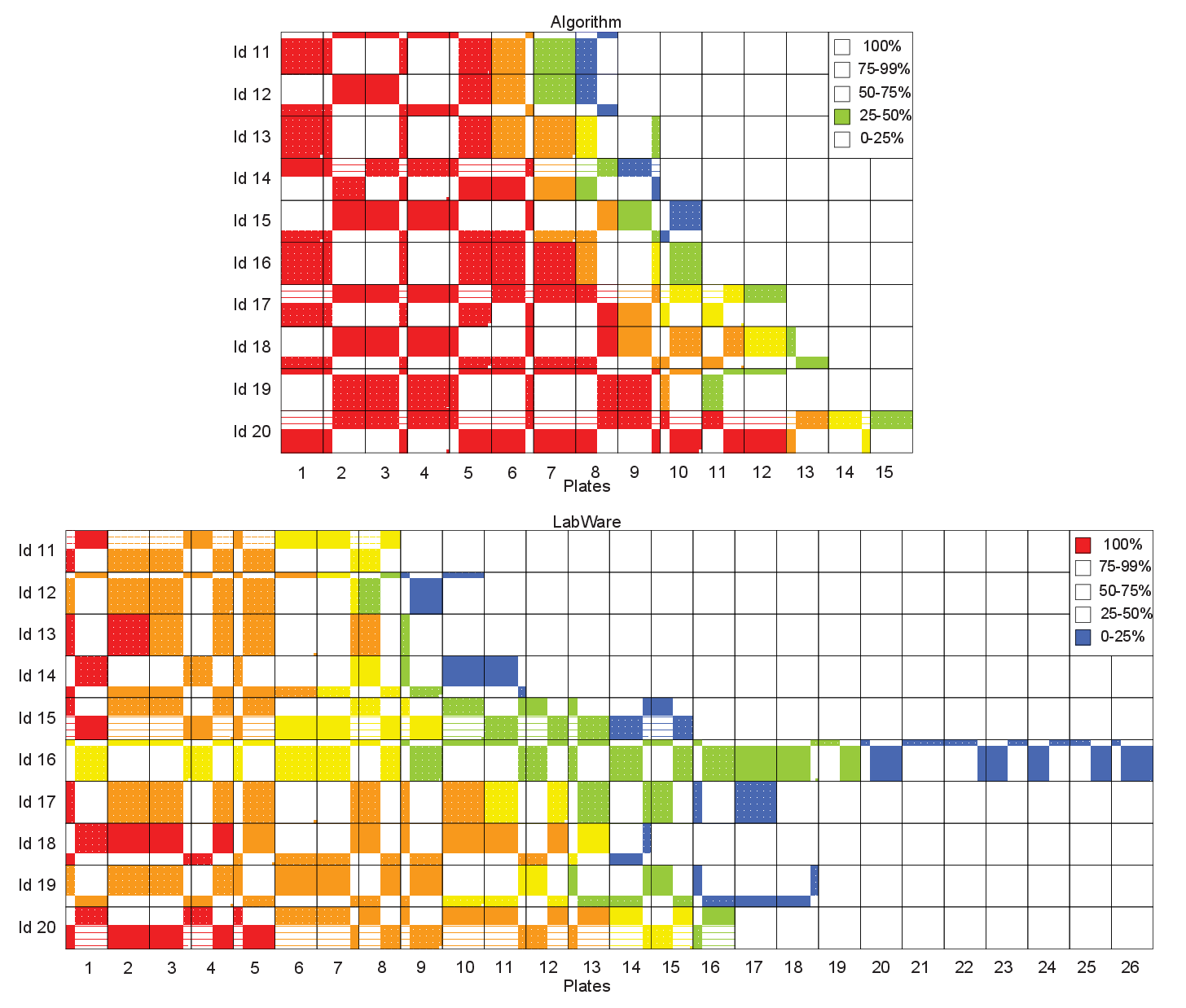}

    \caption{Distribution of the plate occupancy obtained with the algorithm and LabWare in medium instances} \label{mapacolor2}
\end{figure}

As it has been previously mentioned, it is also interesting for the laboratory to have a certain distribution of the plate occupancy rates, so that the first plates are more occupied than the last plates. This would allow new samples to be placed on the most empty plates, which have not yet been processed, as these are the ones that occupy the last positions. Figure \ref{mapacolor1} and Figure \ref{mapacolor2} show the distribution of the plate occupancy with the algorithm and Labware in small and medium instances, respectively.

In the case of both  figures, each row represents the best solution obtained for the corresponding instance by the proposed algorithm and by LabWare. The coloured squares represent the occupied plates and the occupancy level (100\%, [75,100)\%, [50,75)\%, [25,50)\% and [0,25)\%) is indicated by the different colours. It can be clearly seen that both the number of plates used and the number of full plates is worse in the solution obtained by LabWare.

\begin{figure}[h]
    \centering
    \subfigure{\includegraphics[width=10cm]{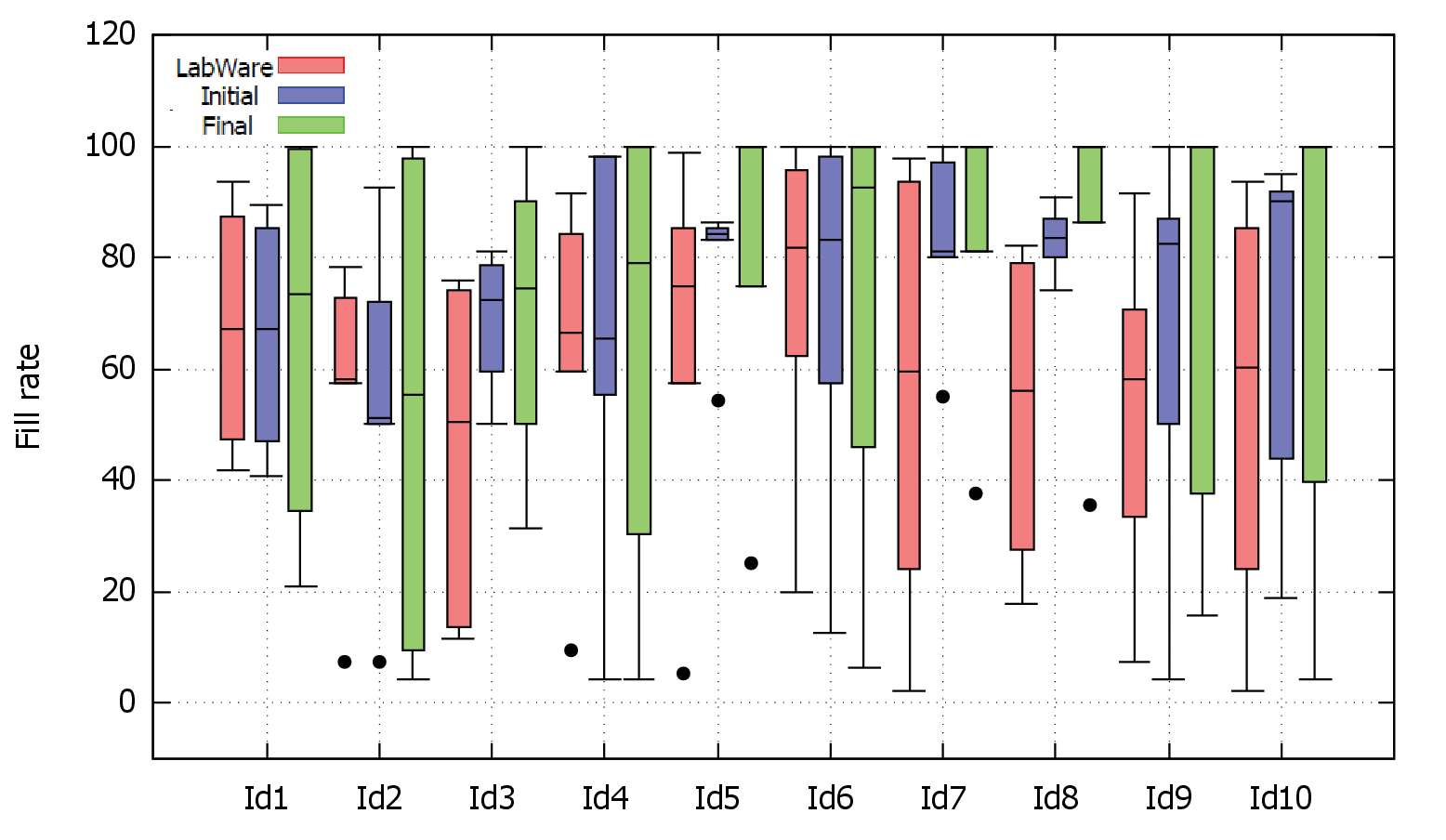}}
    \subfigure{\includegraphics[width=10cm]{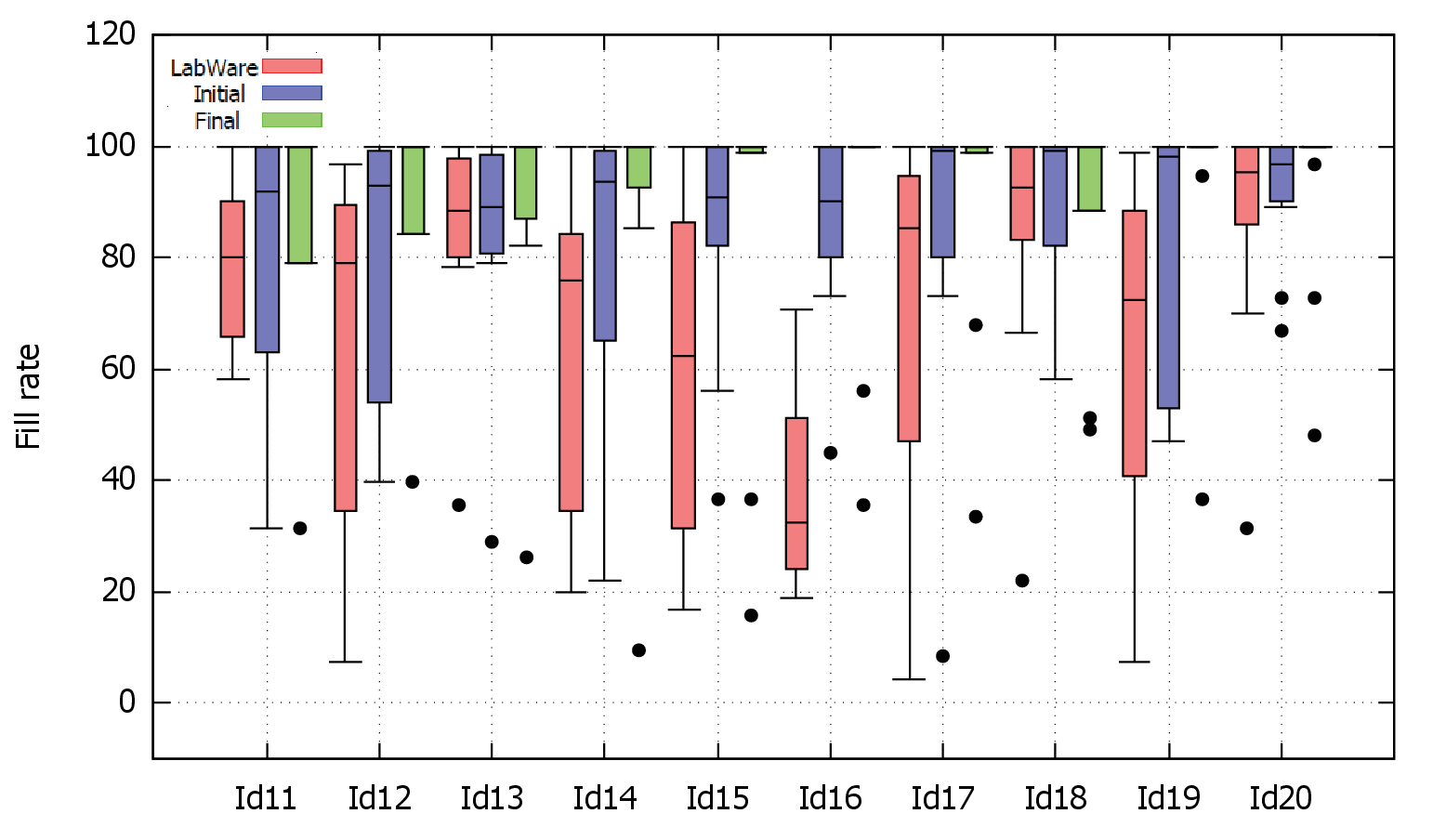}}
    \caption{Comparison of the fill rates that were obtained by LabWare, the initial solution, and the final solution of the algorithm}
    \label{fig:small-med}
\end{figure}

To highlight the efficiency of the heuristic algorithm against the solution that was obtained with LabWare, Figure \ref{fig:small-med} compares, by means of box-plots, the fill rates that were obtained by the initial and final solutions of the algorithm with the ones provided by LabWare for small and medium instances (Files 1-20).
In the case of the medium instances, the solutions of the algorithm yield much higher fill rates than LabWare's solution.

\begin{figure}[htbp]
\centering \subfigure{\includegraphics[scale=0.3]{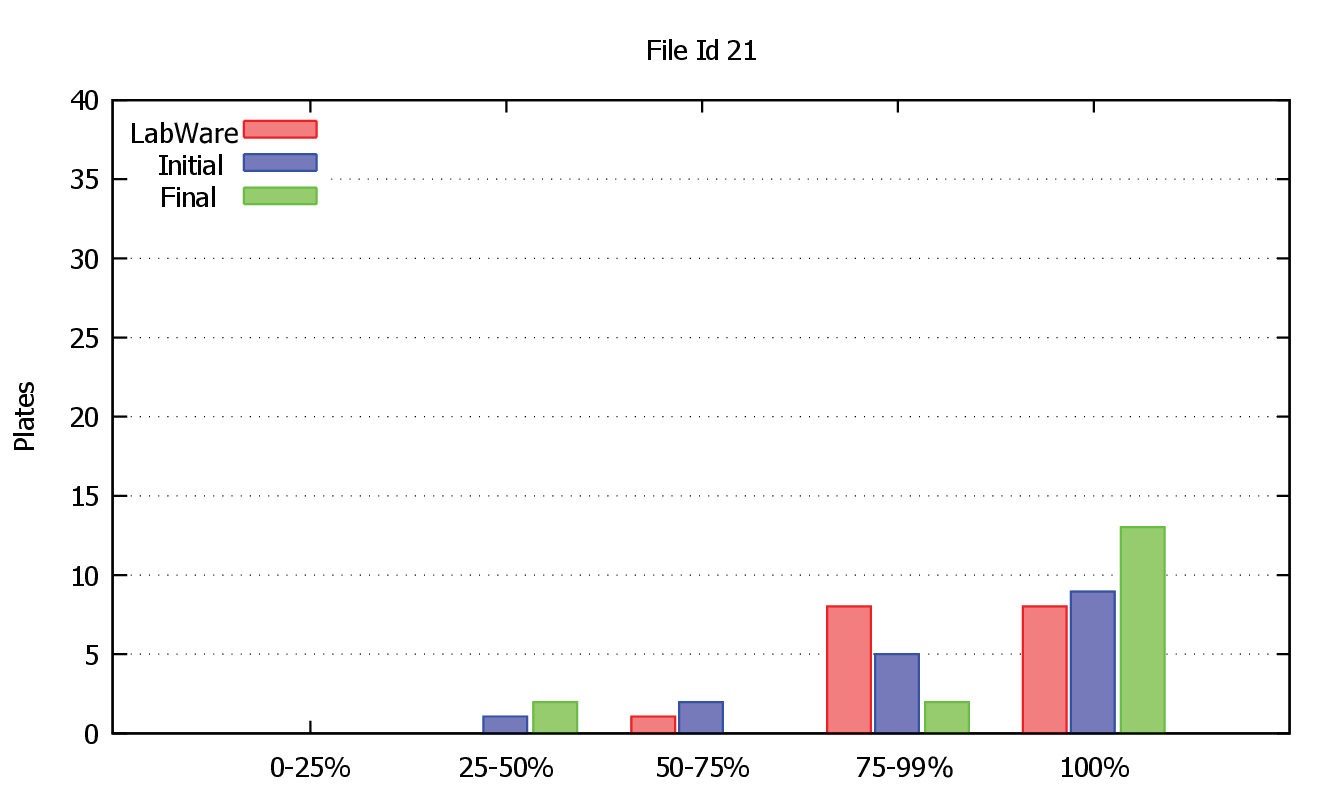}}
\subfigure{\includegraphics[scale=0.3]{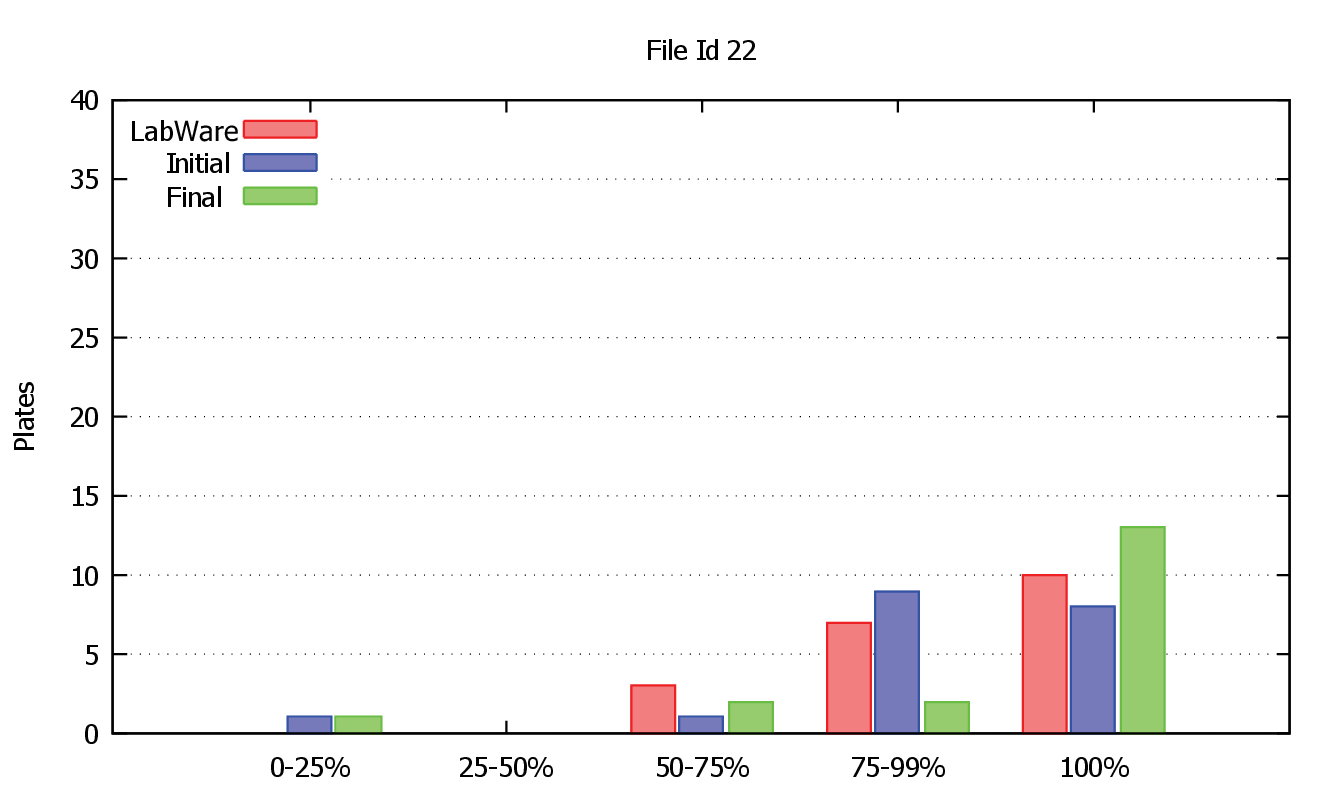}}
\subfigure{\includegraphics[scale=0.3]{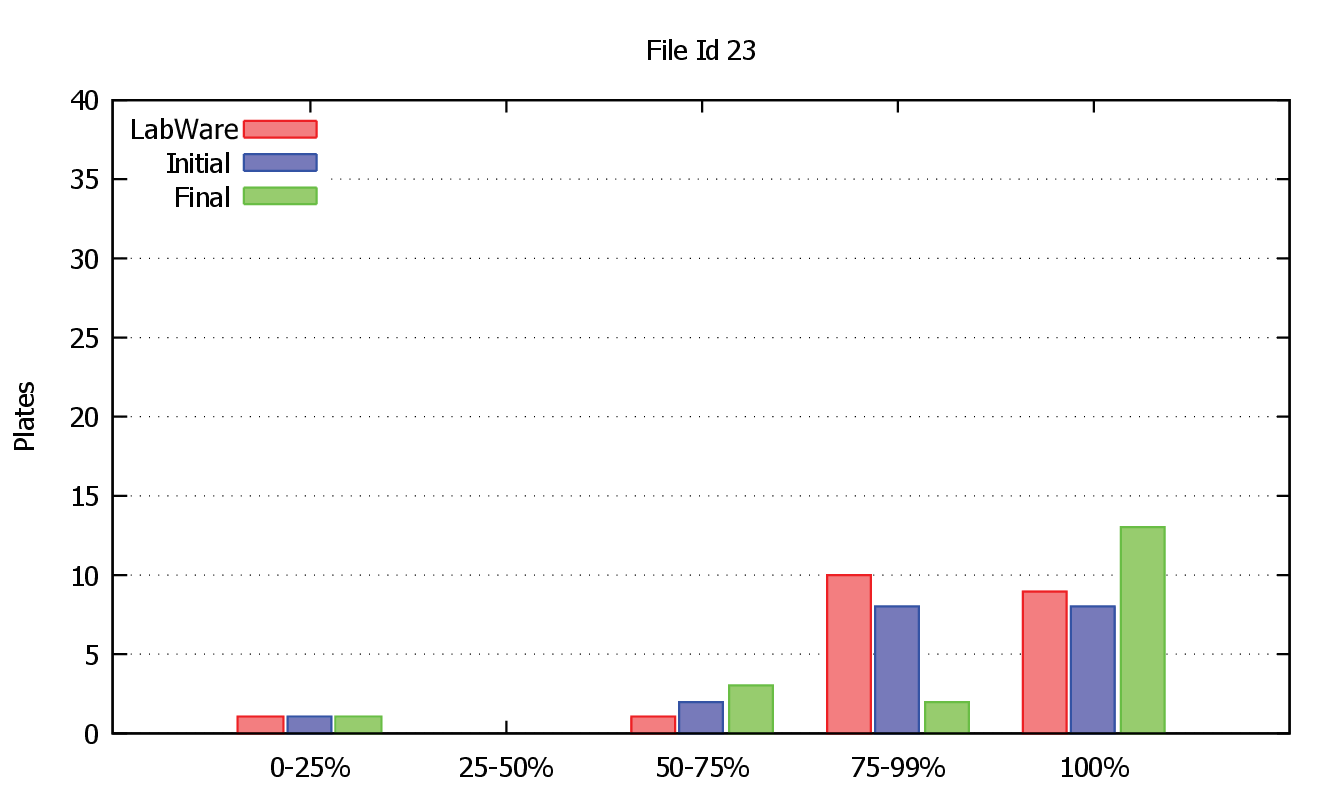}}
\subfigure{\includegraphics[scale=0.3]{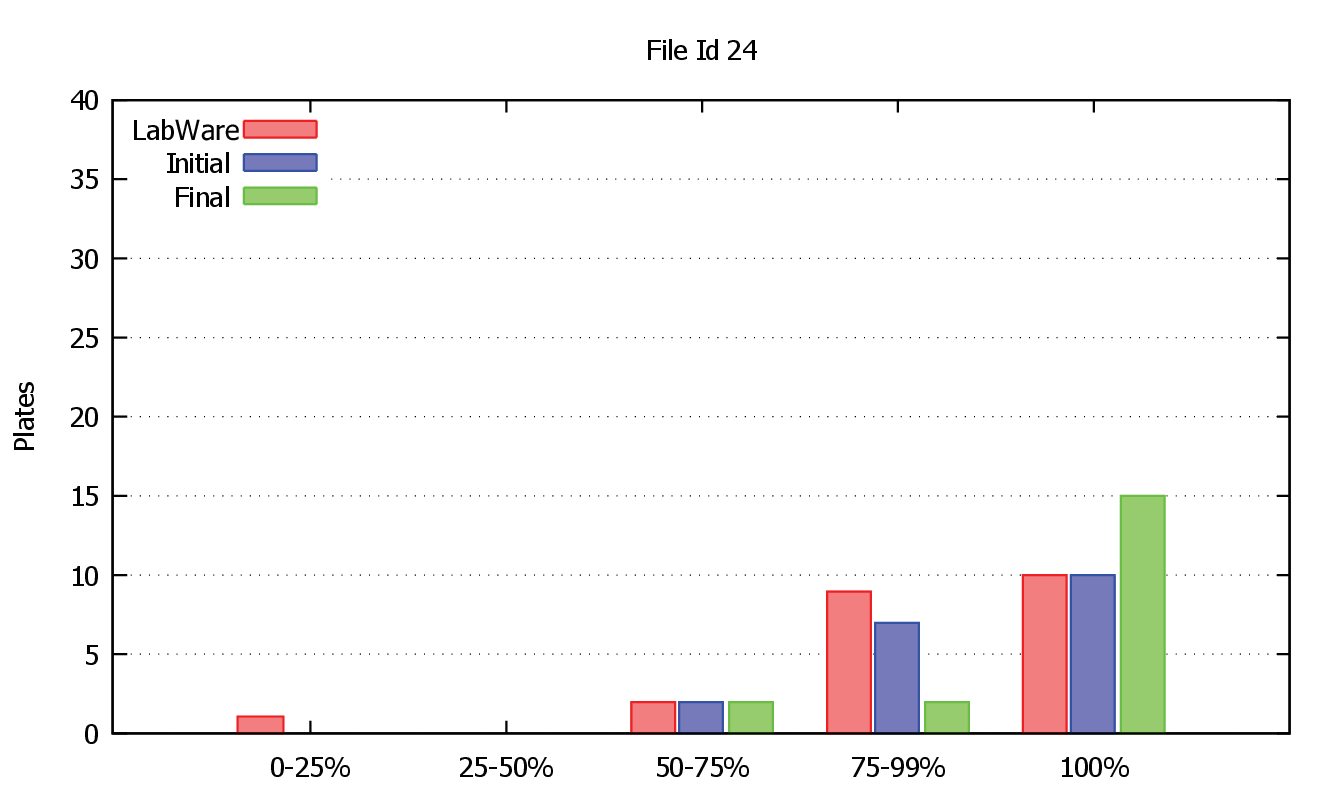}}
\subfigure{\includegraphics[scale=0.3]{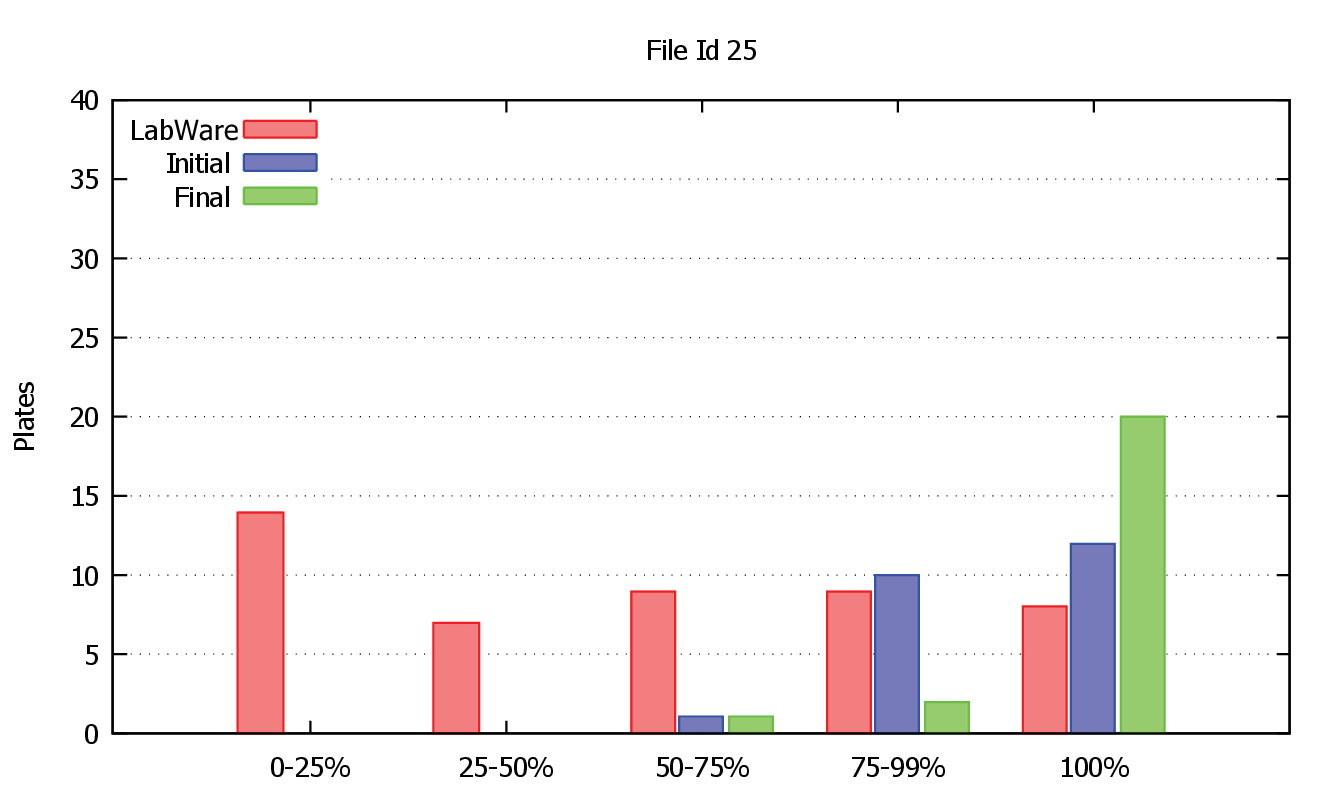}}
\subfigure{\includegraphics[scale=0.3]{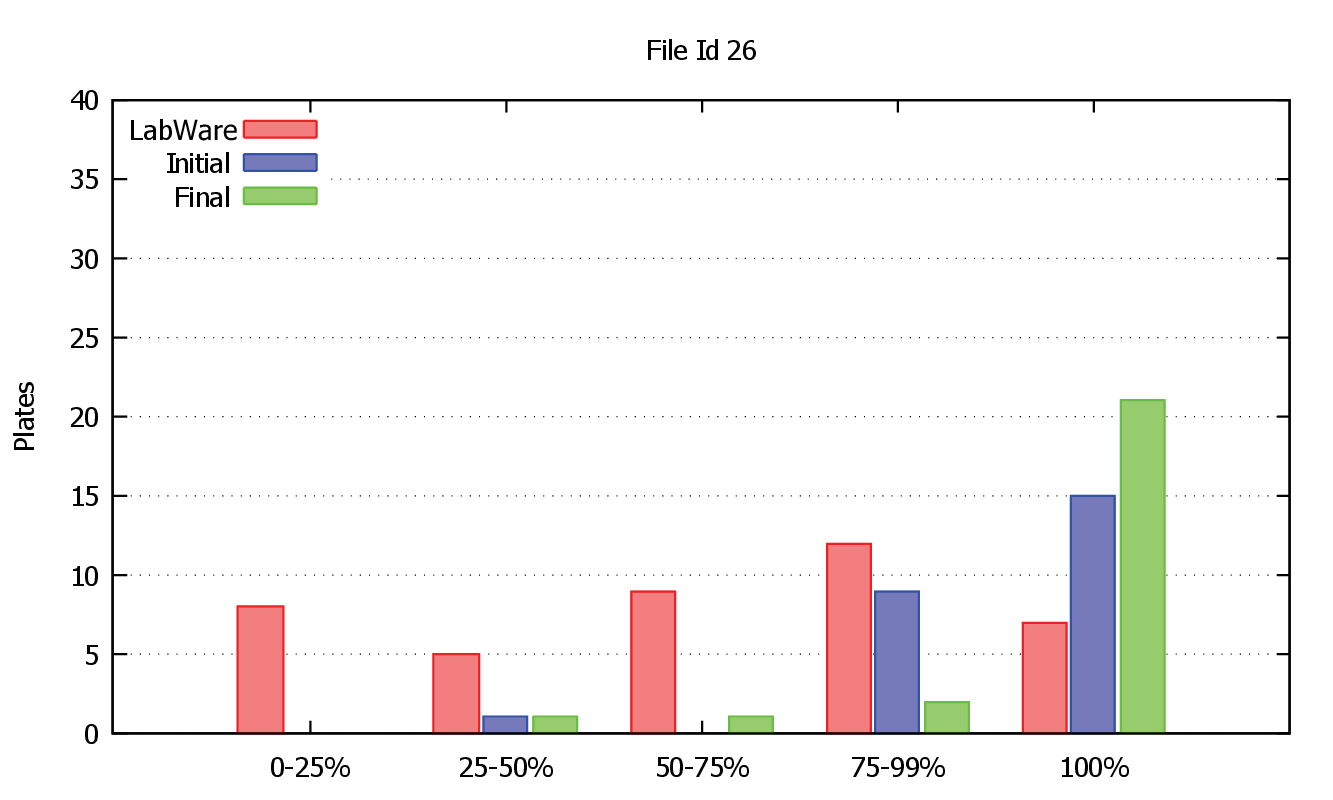}}
\subfigure{\includegraphics[scale=0.3]{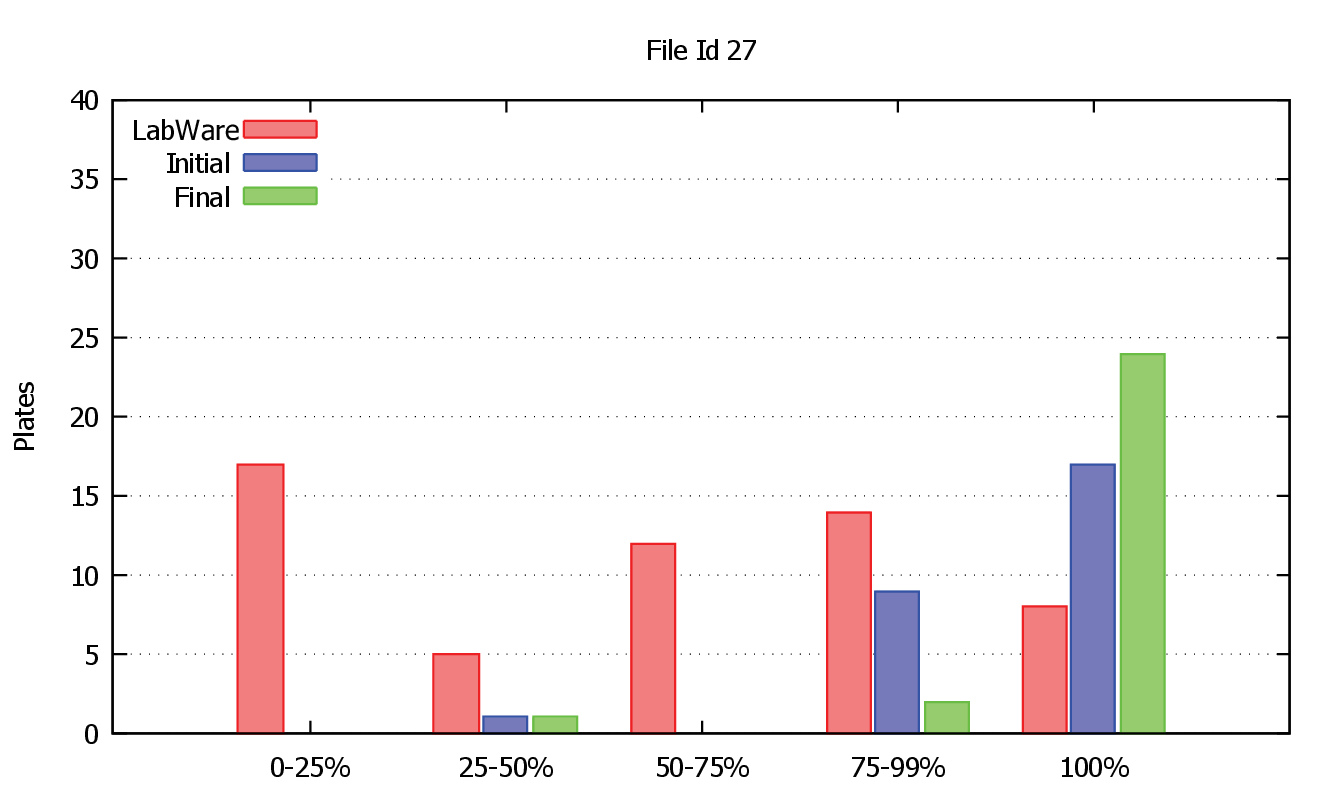}}
\subfigure{\includegraphics[scale=0.3]{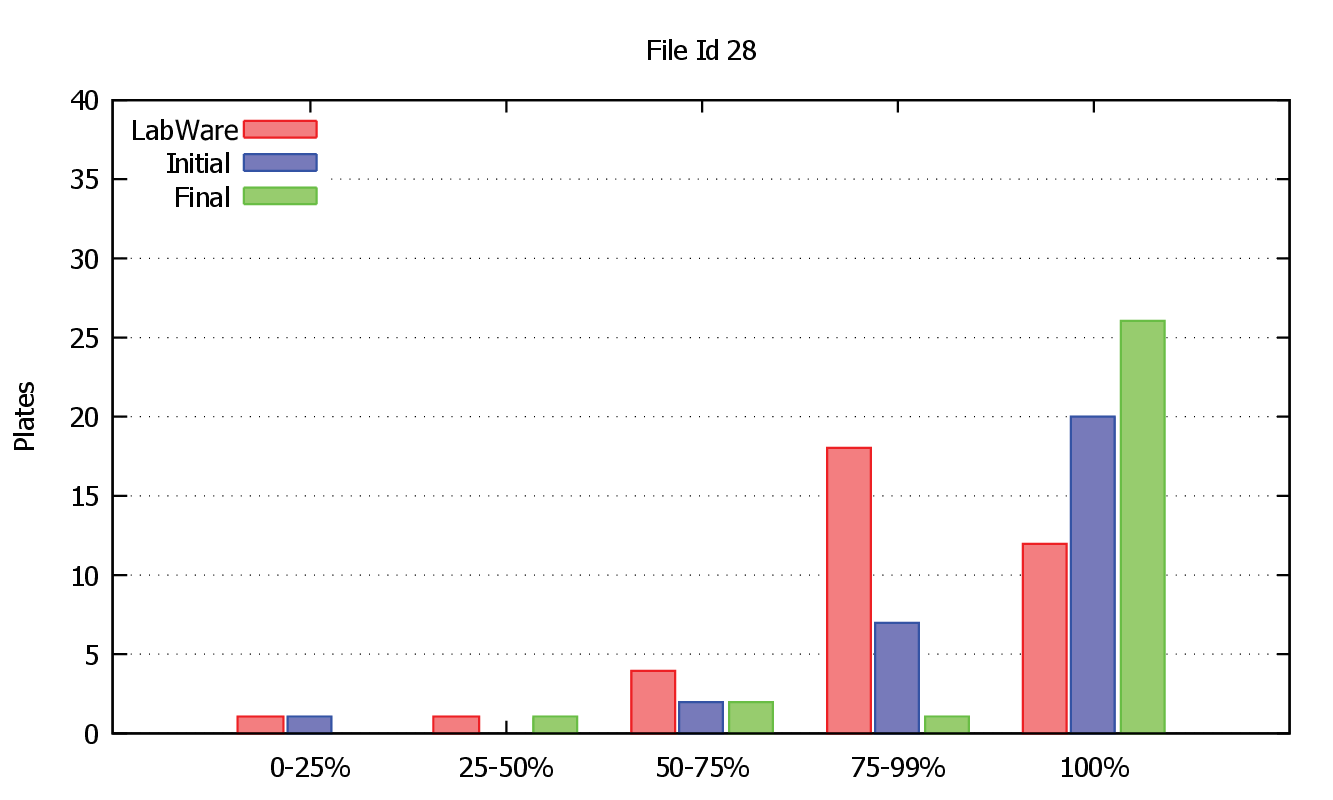}}
\subfigure{\includegraphics[scale=0.3]{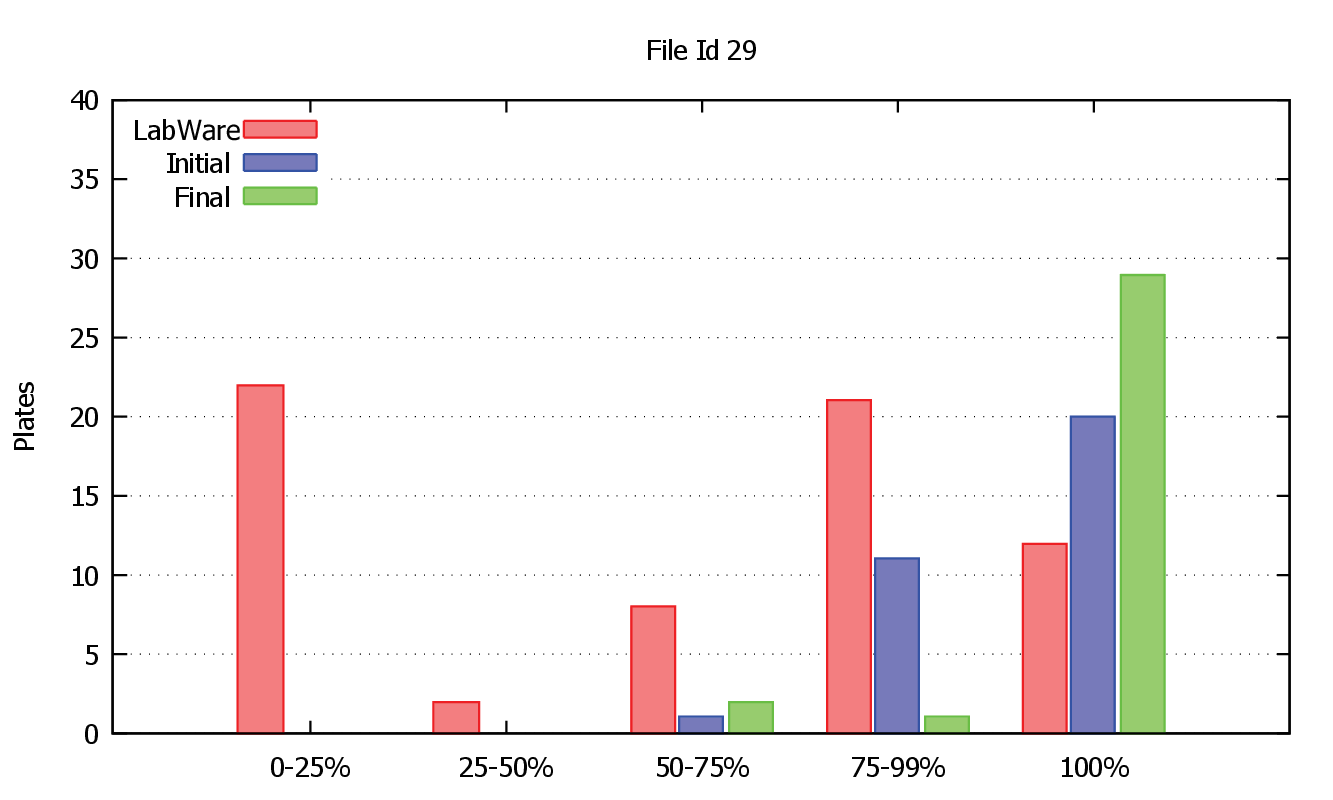}}
\subfigure{\includegraphics[scale=0.3]{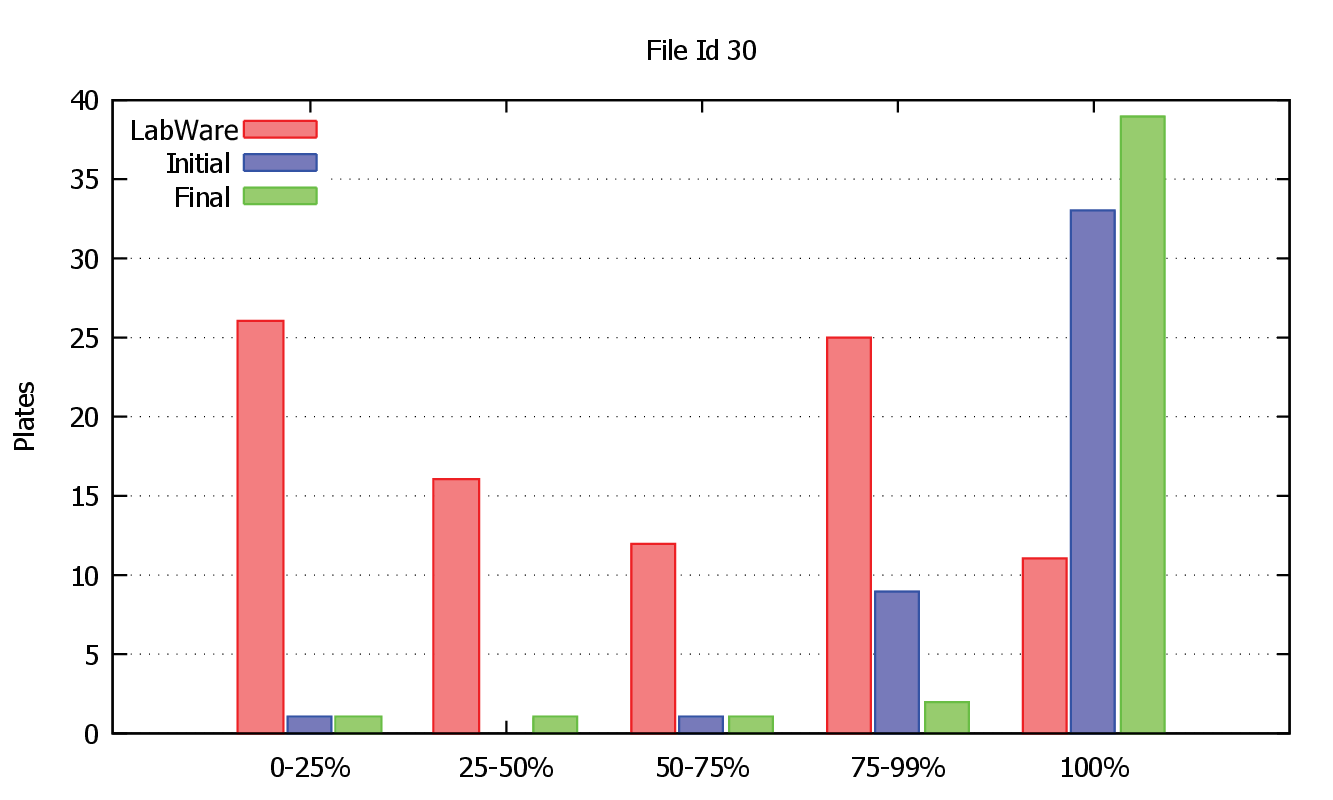}} \caption{Distribution of the plate occupancy obtained with the initial and final solutions of the  algorithm and LabWare in large  instances}
 \label{percentages}
\end{figure}

Finally, Figure \ref{percentages} compares the filling distributions of initial and final solutions that were obtained via the heuristic algorithm with the solution provided by LabWare for large instances (Files 21-30). As a summary of the obtained results, the number of plates that correspond to each range of filling percentages (100, [75,100), [50,75), [25,50) and [0,25)) is displayed for each file. Thus, for instance, in the case of the largest file (Id $30$), the heuristic algorithm finds a scheme that fills $39$ plates out of a total of $44$ plates, while the LabWare solution fills $11$ plates out of a total of $90$.

\section*{Concluding remarks}
In this paper, a real problem that was proposed by Health in Code is studied. This company specializes in genetic diagnosis services for cardiovascular diseases. One of the diagnosis procedures is based on the Sanger method. The Sanger method consists of several phases. In one phase, DNA samples are placed into PCR plates to be processed in thermocyclers. The problem  addressed in this paper focusses on the organization of samples in plates such that the minimum number of plates are used since plate processing is expensive.  Minimization of plate use would also enable Health in Code to offer more competitive prices in the market, as an increasing number of laboratories are dedicated to DNA sequencing.

The scheduling problem considered in this paper is difficult to solve since to organize the samples on the plates, a series of constraints must be satisfied in such a way that this problem differs from other problems that have been studied in the literature.  The laboratory uses 96-well PCR plates and the wells are organized into 8 rows and 12 columns. Each plate is composed of six \textit{strips}. Each strip has 24 wells, which are arranged in 8 rows and 2 columns. According to the characteristics of the thermocyclers, all the wells in the same strip will be processed at the same temperature. In addition, the difference in temperature between two consecutive strips cannot exceed 5 degrees centigrade and one well of the plate should be reserved for the isolated reagent that is associated with a group. In addition, the laboratory faces work sessions in which thousands of samples must be processed.

First, an integer linear programming model was developed. It has been shown that with the ILP model, only small problems can be solved, whereas real problems are burdensome. To overcome this, a heuristic algorithm (based on the simulated annealing philosophy) has been designed. This algorithm obtains satisfactory solutions in short amounts of time and, even in small problems for which the ILP model can be used, provides solutions of similar quality in a much shorter computational time, as the company demands. Moreover, these solutions substantially outperform the solutions that are obtained by the LabWare software, which was used previously at Health in Code. In most cases, it is possible to substantially reduce the number of plates needed for the samples.

The algorithm presented in this paper has been successfully implemented in the laboratories of Health in Code and corresponds to software registration 03/2017/560, which is entitled \textit{"SimPCR: librer\'{\i}a para la optimizaci\'on del proceso de llenado de placas PCR en
    secuenciaci\'on Sanger"}.

\section*{Acknowledgements}

This work has been supported by MINECO: MTM2014-53395-C3-1-P, MINECO: MTM2017-87197-C3-1-P,  Xunta de Galicia/FEDER-UE ERDF: ED431C-2016-015,  Xunta de Galicia/FEDER-UE ERDF: ED431G/01, FEDER-UE  ESF, Xunta de Galicia Conecta Peme-2014: IN852A-2014/9, Xunta de Galicia/FEDER-UE CSI: ED431G/01, Xunta de Galicia/FEDER-UE GRC: ED431C 2017/58, MINECO-CDTI/FEDER-UE CIEN LPS-BIGGER: IDI-20141259, MINECO-CDTI/FEDER-UE INNTERCONECTA uForest: ITC-20161074, MINECO-AEI/FEDER-UE eDSalud: RTC-2016-5143-1, MINECO-AEI/FEDER-UE Datos 4.0: TIN2016-78011-C4-1-R and MINECO-AEI/FEDER-UE ETOME-RDFD3: TIN2015-69951-R.

The interesting and constructive comments made by three anonymous referees are also gratefully acknowledged.


\end{document}